\documentclass[prb,aps,longbibliography]{revtex4-2}
\usepackage{dcolumn}
\usepackage{amsmath}
\usepackage{amsfonts}
\usepackage{amssymb}
\usepackage{calligra}
\usepackage{bm}
\usepackage{graphicx}
\usepackage{float}
\usepackage{subfigure}
\usepackage{stackengine}
\usepackage{natbib}
\usepackage{hyperref}
\usepackage{algorithm}
\usepackage{algpseudocode} 
\usepackage{algorithmicx}  
\usepackage{physics}
\usepackage{grffile}

\begin{document}

\title{
Phase Structure and Machine Learning Identification in One Dimensional Systems with Power Law Correlated Disorder and Long Range Hopping
}

\author{Mohammad Pouranvari}
\affiliation {Department of Physics, Faculty of Basic Sciences, University of Mazandaran, P. O. Box 47416-95447, Babolsar, Iran}


\date{\today}
\begin{abstract}
	We investigate a one-dimensional tight-binding model in which onsite
	potentials $\{\varepsilon_i\}$ exhibit power-law spatial correlations
	(with exponent $\alpha$) and the hopping amplitudes decay as
	$t_{ij}\sim |i-j|^{-\beta}$.  This two-parameter family interpolates
	continuously between short-range Anderson-like disorder, correlated
	disorder with conventional hopping, and long-range hopping models with
	nontrivial delocalization tendencies.  Using large-scale exact
	diagonalization, we construct a comprehensive phase map in the
	$(\alpha,\beta)$ plane by combining spectral statistics, density-of-states
	analysis, and energy-resolved localization indicators such as the
	participation ratio, single-particle entanglement entropy, level-spacing
	ratio $r$, and the ratio of the geometric to arithmetic density of
	states.  From these observables we define phase-indicator functions that
	compactly quantify localization behavior across the spectrum. Our analysis reveals robust mobility edges and multiple regimes of
	spectral coexistence between localized, extended, resonant, and critical
	states.  Finite-size scaling, implemented via an explicit smoothness-based
	cost function, enables extraction of critical exponents and delineation
	of transition lines across the $(\alpha,\beta)$ parameter space.
	To validate and complement these physics-based diagnostics, we employ a
	supervised autoencoder that learns high-level representations of
	eigenstate structure directly from raw features and reliably reproduces
	the phase classification defined by the indicator functions.  Together,
	these approaches provide a coherent and internally consistent picture of
	the spectral transitions driven by correlated disorder and long-range
	hopping, establishing a unified framework for characterizing mobility
	edges in long-range one-dimensional systems.
\end{abstract}

\maketitle

\section{Introduction}
\label{sec:intro}

The interplay of disorder and quantum interference in low-dimensional
systems is a paradigmatic problem in condensed-matter physics\cite{PhysRevA.71.032321, PhysRevB.88.081101, walschaers2016quantum, dobrinevski2011interference, mishra2016constructive}.  In one
dimension, the conventional result is simple and robust: for
short-range models with uncorrelated disorder, all single-particle
eigenstates are exponentially localized in the thermodynamic limit
\cite{Anderson1958,Abrahams1979}.  Departures from this simple picture
arise, however, when one of the basic assumptions is relaxed — for
example, when the potential landscape contains long-range correlations
\cite{deMoura1998,izrailev1999localization, herbut2000commentlocalizationmobilityedge} or when the kinetic energy is
long-ranged \cite{Levitov1990,mirlin1996transition}.  Each of these ingredients
alone can produce atypical spectral and transport behavior in one
dimension; when combined the resulting phenomenology becomes rich and
— in many respects — only partially charted. 	On one hand, tight-binding models with long-range hopping, where the hopping amplitude between sites $i$ and $j$ decays as a power law $|i-j|^{-\beta}$, can host a localization-delocalization transition even in 1D\cite{mirlin1996transition, rodriguez200011}. For sufficiently slow decay ($\beta \le 1$), the hopping is non-local enough to overcome the localizing effects of disorder, leading to extended eigenstates\cite{mirlin1996transition}. Such models have been instrumental in understanding phenomena ranging from transport in disordered systems to the nature of eigenstates in power-law random banded matrices, which often exhibit multifractal properties at the transition point~\cite{mirlin1996transition, perez2019interplay, celardo2016, levitov1999critical, nag2019many, singh2017effect, zhou2003one, qi2021topological, wang2023quantum}.

On the other hand, introducing spatial correlations into the disorder potential can also induce delocalization. Potentials with long-range correlations, such as those generated from fractional Brownian motion, can create extended states and even mobility edges separating localized and delocalized states within the energy spectrum~\cite{deMoura1998,izrailev1999localization}. The intuition is that correlated disorder is "smoother" than white-noise disorder, reducing the effectiveness of random scattering\cite{keen2015crystallography, carpena2002metal, neverov2022correlated, PhysRevLett.107.156601, Pouranvari_2025}.

While these two mechanisms for delocalization have been studied extensively in isolation\cite{mirlin1996transition,deMoura1998,izrailev1999localization}, their interplay remains a rich area of investigation\cite{deng2020mobility, li2017delocalization}. A central question we address is whether, and under what conditions in the $(\alpha,\beta)$ parameter space, there will be a phase transition between delocalized and localized states for the entire spectrum or, more interestingly, whether single-particle mobility edges—energy-dependent transitions between localized and extended eigenstates—appear and remain robust with system size.

The simultaneous presence of long-range correlated disorder and long-range hopping is not merely a theoretical construct but can be realized in several contemporary experimental platforms. For example, trapped-ion chains offer precise control over both onsite potentials and long-range spin-spin interactions, which can be mapped to fermionic hopping with power-law tails\cite{kuhl2000experimental, chavez2021disorder}. Similarly, Rydberg atoms in optical tweezers can simulate disordered Hamiltonians with tunable interactions and disorder correlations\cite{browaeys2020many,chavez2021disorder}. Polar molecules in optical lattices also exhibit dipolar-mediated long-range tunneling and can be subjected to engineered disorder potentials with prescribed spatial correlations\cite{moses2017creation, singh2017effect, ito2003long}. Moreover, synthetic photonic or acoustic lattices with designed refractive-index modulations and long-range coupling elements provide another versatile setting for realizing such models\cite{ozawa2019topological}. These experimental advances make the physics explored in this work directly accessible to laboratory investigation.

It is important to note that in the standard power-law correlated disorder model (the de Moura–Lyra model) with nearest-neighbor hopping, finite-size effects can lead to an apparent transition at $\alpha \approx 2$ in numerical studies of limited system sizes\cite{deMoura1998,izrailev1999localization}, while analytical arguments and more extensive scaling analyzes indicate that the true thermodynamic localization–delocalization transition occurs at $\alpha = 1$\cite{khan2023anomalous, pires2019global, khan2023spectral, khan2023conductance}. This discrepancy highlights the necessity of careful finite-size scaling when diagnosing phase transitions in systems with long-range correlations. In our model, which combines both power-law correlated disorder and long-range hopping, finite-size effects are further compounded by the non-local nature of the kinetic energy. Therefore, throughout this work, we employ systematic finite-size scaling of energy-resolved diagnostics, aided by a smoothness-based cost function to optimize the collapse. Furthermore, we utilize a supervised machine-learning (autoencoder) framework trained directly on finite-size spectral features.

In this work we consider a family of one-dimensional tight-binding
Hamiltonians in which (A) the on-site potentials $\{\varepsilon_i\}$ are constructed by Fourier synthesis with random phases so that the power spectrum scales as $S(k)\propto k^{-\alpha}$; the resulting sequence is mean-subtracted and variance-normalized and is approximately Gaussian for large system size. and (B) the hopping
amplitude between sites decays algebraically as $|i-j|^{-\beta}$
with exponent $\beta$.  This two-parameter family interpolates
between several limits of current interest: (i) the short-range,
uncorrelated Anderson limit ($\beta\to\infty$, small $\alpha$),
(ii) models with long-range correlated potentials but
nearest-neighbour hopping ($\beta\gtrsim 1$, $\alpha \gg 1$), (iii)
long-range hopping chains that have been argued to admit nontrivial
critical behavior and delocalization tendencies for sufficiently
small $\beta$ , and (iv)
long-range hopping with strongly correlated disorder with the
expectation (at first sight) of having extended states across all
spectrum.  Physically, the correlated potential models capture
situations where the disorder arises from a rough substrate or from
long-wavelength structural variations\cite{keen2015crystallography, neverov2022correlated}; long-range hopping arises
naturally in systems with dipolar or Coulomb-mediated tunneling or in
effective single-band descriptions with nonlocal overlap integrals\cite{defenu2021, singh2017effect}.

A central question we address is whether, and under what conditions in
the $(\alpha,\beta)$ parameter space, there will be a phase
transition between delocalized/localized states for the entire
spectrum or more interestingly there will be a single-particle
mobility edges — energy-dependent transitions between localized and
extended eigenstates — appear and remain robust with system size.
Mobility edges in one-dimensional deterministic quasiperiodic systems
are known (for example, in generalizations of the Aubry–André model),
and mobility edges have been reported in various models with
correlated disorder or long-range hopping
\cite{aubry1980analyticity}.  What remains unsettled is
a systematic characterization, using a compact and mutually consistent
set of diagnostics, of how correlated disorder and algebraic hopping
cooperate or compete to produce energy-selective localization.

To address these questions, we use a set of complementary spectral and eigenstate diagnostics. 
We first analyze level statistics through the spacing distribution \(P(s)\) and number variance \(\Sigma^2(L)\), which distinguish Poisson (localized) from Wigner--Dyson (extended) behavior. 
We then employ energy-resolved probes—the participation ratio, single-particle entanglement entropy, and the auxiliary measures \(S_{\mathrm{proxy}}\) and the \(\rho\)-ratio—to characterize eigenstate structure across the band. 
Finally, we define four phase-classification functions \(f_{1}\!-\!f_{4}\), which quantify the fractions of energy bins labelled localized, extended, resonant, or critical, providing a compact summary of spectral mixing for constructing the \((\alpha,\beta)\) phase diagram.

A central emphasis of this work is that energy-resolved analysis is 
essential: mobility edges appear throughout broad regions of the 
$(\alpha,\beta)$ parameter space, and global (spectrum-averaged) 
quantities obscure the coexistence of localized and extended states.  
To quantify transitions we adopt a finite-size scaling ansatz together 
with a smoothness-based cost function, providing an operational route to 
extract critical points and scaling exponents directly from finite-size 
data.  
As an independent cross-check, and to extract high-level structure from 
raw eigenstate features, we train a supervised autoencoder whose 
latent-space representation correlates strongly with the physics-based 
$f$-indicators and cleanly resolves localized, extended, resonant, and 
critical regimes.

Main results of the paper are:

\begin{itemize}
	
	\item We show that the combined action of power-law–correlated onsite potentials (tuned by $\alpha$) and algebraically decaying hopping amplitudes (tuned by $\beta$) generates extended regions in the $(\alpha,\beta)$ parameter space where the spectrum exhibits clear energy-dependent separation between localized and extended states—i.e., robust mobility edges.  
	These boundaries are identified through multiple complementary indicators, including the participation ratio, single-particle entanglement entropy, $r$-statistics, the DOS-derived quantity $\rho$, and level-statistics behavior interpolating between Poisson and Wigner–Dyson limits.  
	Across the full spectrum, we resolve four distinct spectral sectors—localized, extended, resonant, and critical—each associated with well-defined patterns in these observables.
	
	\item We introduce four phase-characterization functions $f_1,\ldots,f_4$, each constructed from physically interpretable combinations of PR, EE, level statistics, and DOS-related indicators.  
	Each $f_i$ evaluates the number of spectral bins exhibiting localized, extended, resonant, or critical phenomenology, providing an operational classification framework that compresses the high-dimensional diagnostic data into compact, robust phase maps.
	
	\item By implementing a finite-size scaling collapse governed by a smoothness-adaptive cost function, we extract transition points and effective scaling exponents for representative cuts through the $(\alpha,\beta)$ plane.  
	Where the collapse is stable, the inferred critical behavior is consistent across independent diagnostics, providing a cross-validated characterization of the mobility-edge transitions.
	
	\item A supervised autoencoder trained on the window-resolved diagnostic vectors reproduces, with high fidelity, the phase labels obtained from the physics-informed $f$-functions.  
	The latent coordinate learned by the model serves as a smooth, interpretable parameter that tracks localized, extended, resonant, and critical behavior, and thereby offers an independent machine-learning confirmation of the phase structure.
	
\end{itemize}
This paper is organized as follows. 
Section~\ref{sec:model-methods} introduces the model, detailing the construction of power-law correlated onsite potentials and the implementation of algebraically decaying long-range hopping together with the required finite-size normalization. 
Section~\ref{sec:diag} defines all diagnostic tools used throughout the work, including level-statistics measures, participation ratios, multifractal dimensions, entanglement-based indicators, the $\rho$-ratio, the finite-size scaling ansatz, and the machine-learning pipeline. 
Section~\ref{sec:results} presents the main numerical results: level-statistics behavior, evidence for mobility edges, classification maps, and representative local diagnostics illustrating the coexistence of localized, extended, resonant, and critical states across the spectrum. 
Finally, Section~\ref{sec:conclusion} summarizes the key findings and outlines several open directions for future research.

\section{Model}
\label{sec:model-methods}

We study a single-band tight-binding model on a one-dimensional chain of length $N$ with open boundary conditions.  The Hamiltonian is
\begin{equation}
	\label{eq:hamiltonian}
	H \;=\; \sum_{i=1}^{N} \varepsilon_i\, c_i^\dagger c_i
	- \sum_{i\neq j} t_{ij}\,\big(c_i^\dagger c_j + \mathrm{h.c.}\big),
\end{equation}
where $c_i^\dagger$ ($c_i$) creates (annihilates) a spinless fermion on site $i$.  The onsite potentials $\{\varepsilon_i\}$ are drawn from a Gaussian ensemble with prescribed power-law spatial correlations (see Sec.~\ref{subsec:correlated-disorder}) and the hopping amplitudes $t_{ij}$ decay algebraically with the intersite separation (Sec.~\ref{subsec:hopping}).

\subsection{Power-law correlated onsite potentials}
\label{subsec:correlated-disorder}

We generate real onsite potentials by superposing Fourier modes with power-law amplitudes following the construction introduced in Ref \cite{deMoura1998},

\begin{equation}
	\varepsilon_j = C_N
	\sum_{k=1}^{N/2} k^{-\alpha/2}
	\cos\!\big(2\pi j k/N + \phi_k\big),
	\label{eq:eps_fourier_real}
\end{equation}
where the phases \(\phi_k\) are independent random variables uniformly distributed on \([0,2\pi)\),
\[
\phi_k \sim \mathcal{U}[0,2\pi),
\]
ensuring that each realization is real-valued and statistically translation invariant.  
The prefactor \(C_N\) fixes the finite-\(N\) normalization of the spectrum (e.g. \(C_N=(2\pi/N)^{(1-\alpha)/2}\)), and each realization is shifted and rescaled to have zero mean and unit variance,
\[
\varepsilon_j \mapsto 
\frac{\varepsilon_j-\langle\varepsilon\rangle}{\sqrt{\langle\varepsilon^2\rangle}}.
\]

\begin{figure}[t]
	\centering
	\includegraphics[width=0.95\linewidth]{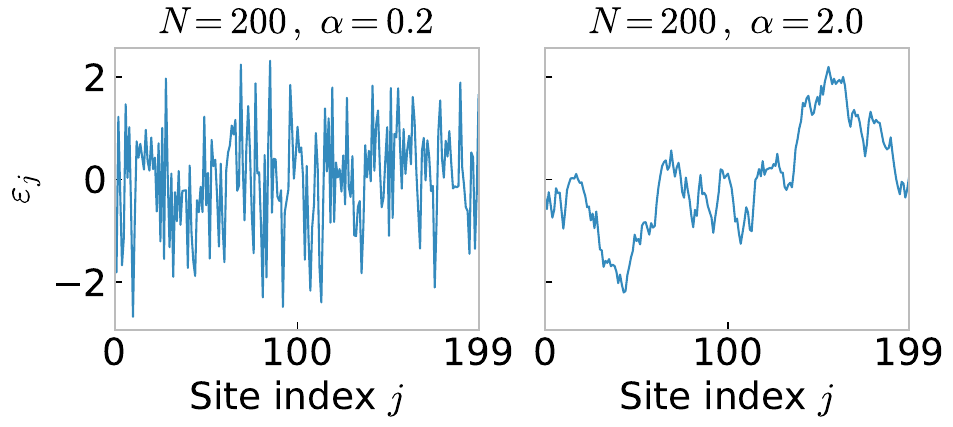}
	  \caption{
Real-space onsite potentials $\varepsilon_j$ for a chain of length $N=200$, obtained by superposing Fourier modes with power-law amplitudes following the construction in Ref \cite{deMoura1998}. (Left panel) $\alpha=0.2$: weakly correlated, rapidly varying (rough) disorder. (Right panel) $\alpha=2$: enhanced low-$k$ weight produces a much smoother, long-range correlated potential. These representative values illustrate the crossover from rough, short-range disorder at small $\alpha$ to smooth, long-range-correlated landscapes at large $\alpha$. Only one disorder realization is plotted without sample averaging.}
	\label{fig:eps}
\end{figure}

This construction yields onsite correlations that decay asymptotically as
\[
\langle\varepsilon_i \varepsilon_{i+r}\rangle \sim r^{-(2\alpha-1)},
\]
so the exponent \(\alpha\) tunes the smoothness of the potential: larger \(\alpha\) produces longer-range correlations, while smaller \(\alpha\) approaches uncorrelated disorder. All onsite potentials used in this work were generated using Eq.~\eqref{eq:eps_fourier_real} (see Figure~\ref{fig:eps})

The expression $\langle \epsilon_i \epsilon_{i+r} \rangle \sim r^{-(2\alpha-1)}$ is an asymptotic result valid for large separations $r \gg 1$ and for $\alpha > 0$. The uncorrelated (white noise) limit is recovered at $\alpha = 0$, where the power spectrum becomes flat, $S(k) \propto \text{constant}$, yielding $\langle \epsilon_i \epsilon_{i+r} \rangle \propto \delta_{r,0}$ and thus no spatial correlations. In this limit, the asymptotic power-law form no longer applies.

\subsection{Algebraic long-range hopping and finite-size normalization}
\label{subsec:hopping}

The matrix elements of the kinetic term are taken to decay algebraically with distance,
\begin{equation}
	\label{eq:hopping}
	t_{ij} \;=\; \frac{t_0}{\mathcal{N}_N(\beta)}\,\frac{1}{|i-j|^{\beta}},
	\qquad i\ne j,
\end{equation}
where $\beta>0$ is the hopping-decay exponent. $t_0$ sets an overall energy scale (we set $t_0\equiv 1$ in all numerical data shown). To ensure a well-defined thermodynamic limit and maintain the extensivity of the kinetic energy, the hopping term is normalized by a system-size-dependent factor $\mathcal{N}_N(\beta)$~\cite{defenu2021, sarkar2024}. This normalization factor is chosen as:
\[
\mathcal{N}_N(\beta) =
\begin{cases}
	\frac{ 2^{\beta-1}}{1-\beta}N^{1 - \beta} & \text{for } \beta < 1, \\
	\log N & \text{for } \beta = 1, \\
	\zeta(\beta) & \text{for } \beta > 1,
\end{cases}
\]
where $\zeta(\beta)$ is the Riemann zeta function. In our numerical calculations for $\beta > 1$, we approximate $\zeta(\beta) \approx 1/(\beta - 1) + \gamma$, where $\gamma \approx 0.5772$ is the Euler–Mascheroni constant.

\section{Observables and Diagnostics}
\label{sec:diag}

To characterize the different phases of the model, we employ several well-established numerical diagnostics. All spectral and eigenstate-based diagnostics in this work are obtained by exact diagonalization of the single-particle Hamiltonian (Eq.\ref{eq:hamiltonian}) for a set of system sizes $N$ (see figure captions for the specific sizes and number of disorder realizations used for each panel).  For energy-resolved quantities we typically group eigenstates into energy windows (bins) and report averages and variances across disorder realizations within a chosen window.  All of these are computed per eigenstate and then averaged (or aggregated) in energy windows as needed.

\subsection{Level Statistics.}
The statistical properties of the energy eigenvalues provide a powerful probe of localization. We analyze the distribution of adjacent level spacings, $s_n = E_{n+1} - E_n$, after an unfolding procedure to ensure a uniform mean level density. For localized states, the eigenvalues are uncorrelated, leading to a Poisson distribution, $P(s) = e^{-s}$. For delocalized, chaotic states described by random matrix theory, level repulsion leads to a Wigner-Dyson (GOE) distribution, $P(s) = (\pi s/2) e^{-\pi s^2/4}$. We also compute the average ratio of adjacent level spacings, $\langle r_n \rangle = \langle \min(s_n, s_{n-1}) / \max(s_n, s_{n-1}) \rangle$, which avoids the need for unfolding. Its value distinguishes between the Poisson limit ($\langle r_n \rangle \approx 0.386$) and the GOE limit ($\langle r_n \rangle \approx 0.531$)~\cite{oganesyan2007}.

From the spacings we form the normalized spacing distribution $P(s)$ with $s\equiv\delta/\langle\delta\rangle$ and the (ensemble-averaged) number variance
\begin{equation}
	\label{eq:number_variance}
	\Sigma^2(L) \;=\; \big\langle\big( N_L - \langle N_L\rangle \big)^2\big\rangle,
\end{equation}
where $N_L$ is the number of eigenvalues in an interval of length $L$ (in unfolded units) and the averages are taken over spectral location and disorder realizations.  

\subsection{Participation ratio}

For a normalized single-particle eigenstate $\psi_n(i)$ (with eigenstate index $n$), the participation ratio is defined as
\begin{equation}
	\label{eq:pr_def}
	\mathrm{PR}_n = \frac{1}{\sum_{i=1}^{N} |\psi_n(i)|^{4}},
\end{equation}
and serves as a measure of the number of sites that significantly contribute to the eigenstate. In the thermodynamic limit, localized states exhibit $\mathrm{PR}_n = \mathcal{O}(1)$, whereas fully extended states scale as $\mathrm{PR}_n = \mathcal{O}(N)$. 

For convenience, we also consider the normalized participation ratio $\mathrm{PR}_n/N$, which vanishes in the localized regime and approaches unity for delocalized states.

\subsection{Generalized multifractal dimensions \(D(q)\)}

To probe possible multifractality of single-particle eigenstates, we
compute the generalized participation moments
\begin{equation}
	P_q(n) = \sum_{i=1}^{N} |\psi_n(i)|^{2q},
\end{equation}
from which the multifractal exponents are defined via the standard
scaling relation
\begin{equation}
	D_q(n) = -\frac{1}{(q-1)\,\log N}\,\log P_q(n).
\end{equation}
In the thermodynamic limit, localized states yield $D_q \rightarrow 0$,
whereas fully extended states satisfy $D_q \rightarrow 1$ for all $q$.
Intermediate values $0 < D_q < 1$ signal multifractal (critical)
eigenstates.  Throughout this work we evaluate $D_q$ for a discrete set
of $q$ values to identify critical spectral windows and to supplement
the other localization diagnostics.

\subsection{Entanglement entropy}
\label{subsec:spee}

To characterize the spatial structure of single-particle eigenstates, we employ two entanglement-based diagnostics.

\paragraph{(i) Single-particle entanglement entropy.}

Here the entanglement entropy (EE) is defined for \emph{single-particle} wavefunctions and should not be confused with the entanglement of many-body eigenstates. For a normalized single-particle state
\begin{equation}
	\ket{\psi} = \sum_{i=1}^{N} \psi_i\, \ket{1_i},
\end{equation}
we bipartition the system into two equal halves, $A$ and $B$. The probability of finding the particle in subsystem $A$ is
\begin{equation}
	p_A = \sum_{i \in A} |\psi_i|^2,
	\qquad
	p_B = 1 - p_A.
\end{equation}
Tracing out subsystem $B$ yields a reduced density matrix of rank~2,
\begin{equation}
	\rho_A = p_A\, \ket{1_A}\bra{1_A} + p_B\, \ket{0_A}\bra{0_A},
\end{equation}
from which the single-particle entanglement entropy is obtained:
\begin{equation}
	EE = -\left(p_A \ln p_A + p_B \ln p_B\right).
\end{equation}
This entropy is maximal for extended states, where $p_A \approx p_B \approx 1/2$, and is strongly suppressed for localized eigenstates, for which $p_A$ is typically close to either 0 or 1.

\paragraph{(ii) Entanglement proxy (\texttt{sproxy}).}

We define the one-sided Shannon-type proxy
\begin{equation}
	s_{\mathrm{proxy}}^{(n)} = - \sum_{i\in A} w_i^{(n)} \ln w_i^{(n)},
	\qquad
	w_i^{(n)} = |\psi_n(i)|^2,
\end{equation}
which measures how the probability weight is distributed {\em within} subsystem $A$.  Unlike the single-particle EE, which depends only on the total weights $p_A$ and $p_B$ and can therefore be large for a state localized near the bipartition (when $p_A\approx p_B\approx 1/2$), $s_{\mathrm{proxy}}$ is small whenever the amplitude in $A$ is concentrated on a few sites.  Consequently, comparing $EE$ and $s_{\mathrm{proxy}}$ allows us to distinguish truly extended states (both $EE$ and $s_{\mathrm{proxy}}$ large) from localized states whose support happens to straddle the cut (large $EE$ but small $s_{\mathrm{proxy}}$).

\subsection{\texorpdfstring{$\rho$}{rho}-ratio}
\label{subsec:rhoratio}

To quantify the degree of spectral mixing within a fixed energy window,
we employ the $\rho$-ratio, a spectral diagnostic constructed directly
from the local density of states (LDOS).  
For an eigenenergy $E_n$ we define the LDOS at site $i$ as 
$\rho_i(E_n)=|\psi_n(i)|^2$.  From this object we construct two coarse-grained 
spectral measures evaluated within a given energy bin: the 
\emph{average} LDOS
\begin{equation}
	\rho_{\mathrm{avg}} = \frac{1}{N_{\mathrm{counted}}}\sum_{i}\rho_i(E_n)
\end{equation}
and the \emph{typical} LDOS,
\begin{equation}
	\rho_{\mathrm{typ}} = 
	\exp\!\left[
	\frac{1}{N_{\mathrm{counted}}}\sum_{i}\ln(\rho_i(E_n))
	\right].
\end{equation}

The $\rho$-ratio is then defined as
\begin{equation}
	R_\rho = \frac{\rho_{\mathrm{typ}}}{\rho_{\mathrm{avg}}}.
\end{equation}
This quantity behaves analogously to the conventional IPR/IPR$_{\mathrm{typ}}$
ratio: it is close to unity when LDOS values are narrowly distributed
(as in extended or ergodic regimes) and strongly suppressed when the 
distribution is broad or log-normal (as in localized regimes).

All observables above can be made energy-resolved by averaging over eigenstates whose energies fall within a chosen energy bin centered at $E$.  Mobility edges (energy-dependent transitions between localized and extended states) are identified by consistent changes in multiple energy-resolved observables.

\subsection{Finite-size scaling ansatz and cost function}
\label{subsec:scaling}

To quantify transitions we adopt the finite-size scaling ansatz\cite{PhysRevB.102.064207},
\begin{equation}
	\label{eq:scaling_ansatz}
	\frac{F(N,t)}{N^{\gamma_1}} \;=\; \mathcal{F}\!\big( N^{\gamma_2}\,(t-t_c) \big),
\end{equation}
where $F(N,t)$ is a suitably chosen observable (for instance a chosen $f_i$), $t$ is the tuning parameter (e.g., $\beta$ or $\alpha$), and $(t_c,\gamma_1,\gamma_2)$ are the critical parameters to be optimized.  In practice we discretize $t$ on a grid and seek the values of $t_c,\gamma_1,\gamma_2$ that minimize a cost function that measures deviations from a single, smooth scaling curve across system sizes.

The cost function used in the paper is a smoothness regularizer that penalizes spurious oscillatory structure.  The smoothness cost used in constructing classification maps is
\begin{equation}
	\label{eq:Csmooth}
	\mathcal{C}
	\;=\;
	\frac{\sum_{i=1}^{N_p-1} \big|Q_{i+1}-Q_i\big|}{\max_i Q_i - \min_i Q_i} \;-\; 1,
\end{equation}
where $\{Q_i\}_{i=1}^{N_p}$ is the sequence of diagnostic values (for example, a monotonic sequence of $f$-values) sampled along a path in parameter space.   minimization proceeds with standard nonlinear optimization routines over the three parameters $(t_c,\gamma_1,\gamma_2)$.

\subsection{Machine-learning pipeline}
\label{subsec:ml}

To complement and independently validate the physics-based diagnostics, we employ a supervised autoencoder (SAE) as a data-driven probe of the emergent phase structure. The objective is not black-box classification, but to assess whether the numerical observables intrinsically encode a low-dimensional organization consistent with the physically defined phase fractions $f_j$.

Each eigenstate (or energy-resolved ensemble) is represented by a compact feature vector built from selected spectral, participation, and entanglement-related observables introduced in Sec.~\ref{sec:results}. The SAE consists of an encoder that maps these inputs to a low-dimensional latent space, a decoder that reconstructs the original features, and a lightweight classifier attached to the latent layer. Training is performed with a composite loss,
\begin{equation}
	\mathcal{L} = \mathcal{L}_{\mathrm{rec}} + \mu\,\mathcal{L}_{\mathrm{class}},
\end{equation}
where $\mathcal{L}_{\mathrm{rec}}$ is a mean-squared reconstruction loss and $\mathcal{L}_{\mathrm{class}}$ is a cross-entropy loss based on coarse-grained phase labels (localized, extended, resonant, and critical) inferred from $f_j(\alpha,\beta)$. The supervision is deliberately weak, acting only to regularize and sharpen the latent representation.

Training proceeds via Adam-type optimization with early stopping based on a validation set. The primary output of the ML pipeline is the geometry of the latent space: if the SAE autonomously organizes the data into manifolds aligned with the physical phases and correlates with $f_j$, this provides an unbiased, internal consistency check of the phase structure discussed in the Results section.

\subsection{Implementation notes and reproducibility}
All numerical spectra were generated using a custom C++ exact-diagonalization
code employing standard LAPACK routines for dense eigensolvers.  For each
pair of parameters $(\alpha,\beta)$ we diagonalized systems up to
$N=512$ sites and accumulated multiple disorder realizations
(the precise number of realizations used for each figure is stated in its caption).
Raw eigenvalues and eigenvectors were stored in text form and subsequently
parsed and analyzed using Python.

Energy-resolved observables (PR, single-particle entanglement entropy,
$s_{\mathrm{proxy}}$, and the density-of-states quantities used for
$\rho_{\mathrm{typ}}$ and $\rho_{\mathrm{ave}}$) were evaluated on the full
spectrum, then coarse-grained within energy windows of width $W=2$ and
binned using $N_{\mathrm{bin}}'=40$ bins unless stated otherwise.  For
windowing-based analyzes, realizations that did not contain sufficient
levels inside the window were automatically discarded.  All global (spectrum-
averaged) quantities were computed only after disorder averaging across the
surviving realizations.

Level-spacing distributions and number-variance data were computed after
standard unfolding using polynomial fits of the integrated density of
states.  All finite-size–scaling collapses were performed using the explicit
cost function described in Sec.~\ref{sec:model-methods}, with systematic
variation of collapse windows and smoothness penalties to monitor stability.

The supervised autoencoder was trained on per-$(\alpha,\beta)$ feature
vectors containing the coarse-grained values of
$\{r,\,\rho_{\mathrm{typ}}/\rho_{\mathrm{ave}},\,\mathrm{PR}_{\mathrm{norm}},
\,s_{\mathrm{proxy}}\}$ across all bins.  Training used an 80/20 split,
early-stopping with patience of 5 epochs, and a fixed latent dimension of
four to match the number of physics-based $f$-indicators.  All ML analyzes
were performed on CPU in standard PyTorch without GPU acceleration.

The autoencoder was trained on feature vectors from a $20 \times 20$ grid of $(\alpha,\beta)$ points (400 points total), each aggregated over 100 disorder realizations. After an 80–20 split, we used 320 samples for training and 80 for validation. To prevent overfitting, we employed early stopping (patience 5), a lightweight architecture (single hidden layer of width 32, latent dimension 4), and a covariance penalty term. Monitoring of training and validation losses showed no significant gap. Additional tests with varying training set sizes (200, 300, 400 samples) indicated that the latent representation saturates beyond $\sim$300 samples. Experiments with deeper networks (2–3 hidden layers) did not improve the correlation with physics-based phase fractions, confirming that the simple architecture adequately captures the essential manifold (see Fig. \ref{fig:latent_vs_f}).

\section*{Code availability}
The custom numerical codes used to generate the results and figures in this study are publicly archived and openly accessible via Zenodo at:
\href{https://doi.org/10.5281/zenodo.19384535}{\texttt{https://doi.org/10.5281/zenodo.19384535}}

\section{results}
\label{sec:results}
In this section, we first study the general picture we can obtain from the adjacent level distribution. Then we look into the entire spectrum
\subsection{Level statistics: spacing distribution and number variance}
\label{subsec:level-stats}

To elucidate the nature of eigenstate statistics and spectral correlations, we compute the level spacing distribution $ P(s) $ and the number variance $ \Sigma^2(L) $ for various representative points in the $(\alpha, \beta)$ parameter space. All data is obtained from spectra unfolded via local polynomial fitting, averaged over multiple disorder realizations. We compare the numerically obtained $ P(s) $ and $ \Sigma^2(L) $ against several theoretical benchmarks: the Poisson distribution, the Wigner-Dyson (WD) distribution, the semi-Poisson form, and the Brody distribution parameterized by the interpolation exponent $ b $.

\begin{figure}
	\centering
	\begin{subfigure}{}%
		\includegraphics[width=0.47\textwidth]{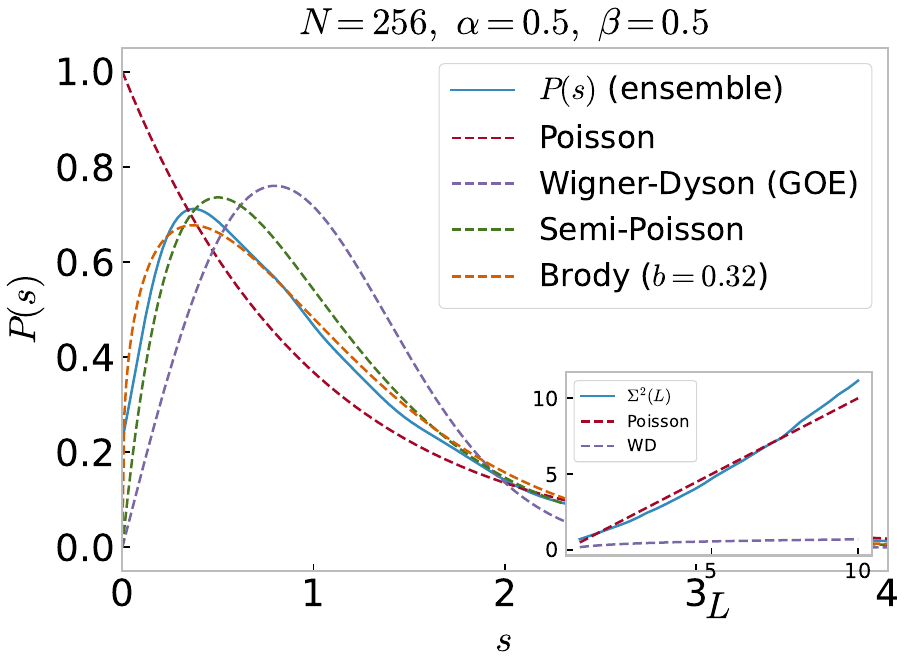}
	\end{subfigure}%
	\begin{subfigure}{}%
		\includegraphics[width=0.47\textwidth]{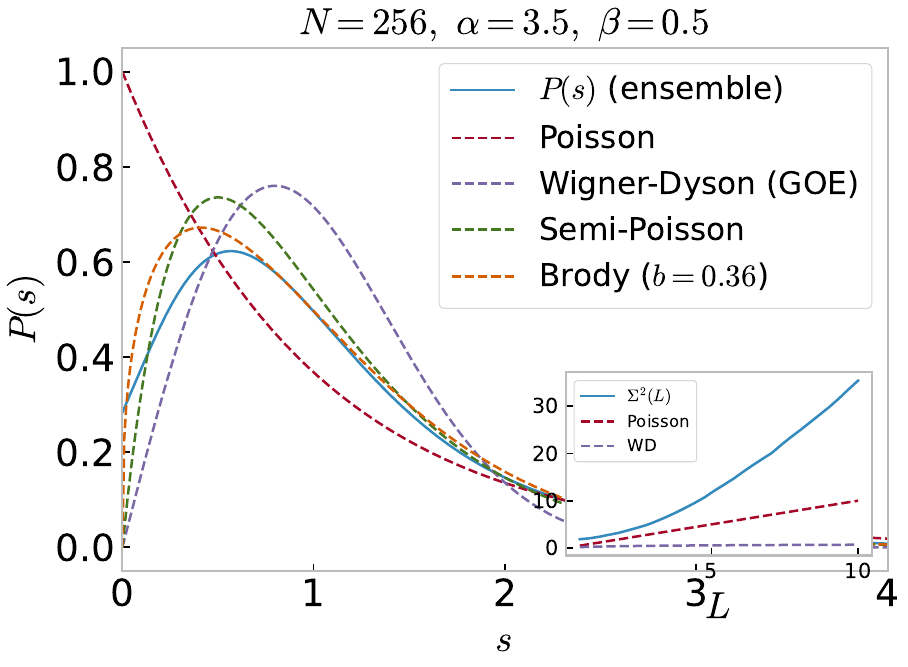}
	\end{subfigure}
	\begin{subfigure}{}%
		\includegraphics[width=0.47\textwidth]{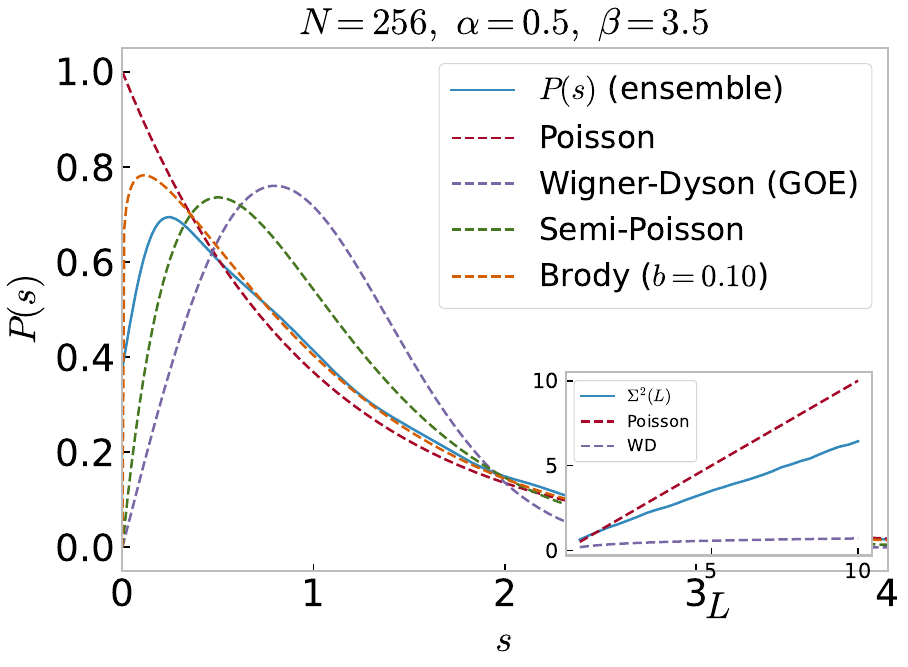}
	\end{subfigure}%
	\begin{subfigure}{}%
		\includegraphics[width=0.47\textwidth]{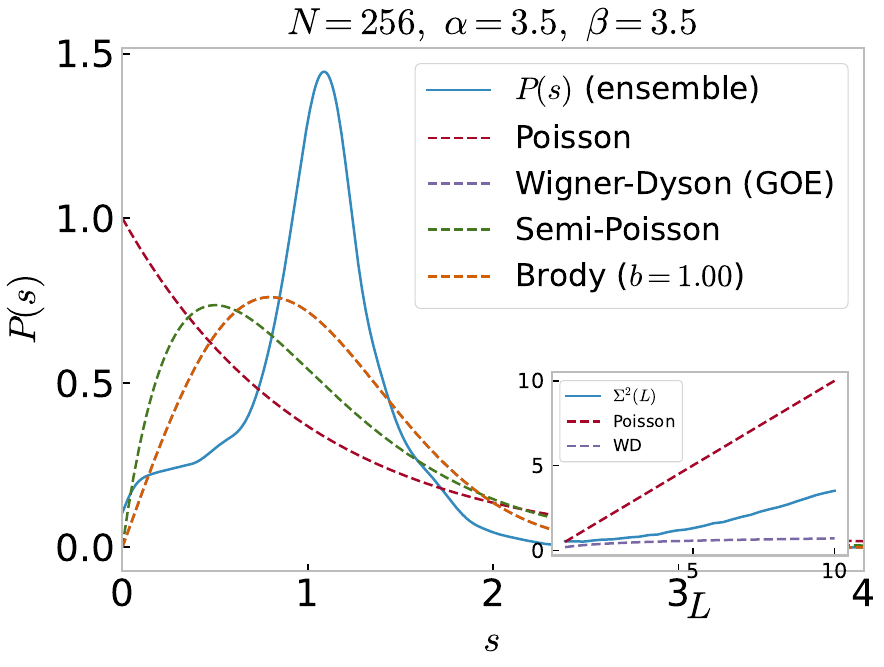}
	\end{subfigure}
	\caption{Level–spacing distributions $P(s)$ and number variances $\Sigma^{2}(L)$ are shown for four corner points in the $(\alpha, \beta)$ parameter space at system size $N = 256$. Each panel compares the numerical spacing distribution with standard theoretical benchmarks: Poisson, Wigner–Dyson (WD), semi–Poisson, and Brody forms. The insets display $\Sigma^{2}(L)$, which further substantiates the evolution from localized to delocalized spectral statistics as functions of the disorder–correlation exponent $\alpha$ and the hopping–range exponent $\beta$. For each parameter set, the results are averaged over $100$ disorder realizations.}
	\label{fig:spacing}
\end{figure}

Figure~\ref{fig:spacing} presents four representative cases corresponding to the corners of the parameter space: 
$(\alpha, \beta) = (0.5, 0.5)$, $(0.5, 3.5)$, $(3.5, 0.5)$, and $(3.5, 3.5)$. These selections are indicative of regimes with weak and strong correlations in both short and long range hopping.

In the weak‐correlation, long‐range hopping regime ($\alpha = 0.5$, $\beta = 0.5$), the level–spacing distribution exhibits a pronounced deviation from the Wigner–Dyson form and falls between the semi–Poisson and Brody distributions (with a fitted Brody parameter $b \approx 0.23$), signaling intermediate spectral statistics. For the same interaction exponent $\alpha$ but with strongly decaying hopping amplitudes ($\beta = 3.5$), the resulting spectrum again lies in neither the Poisson nor the Wigner–Dyson limits, further confirming the presence of intermediate statistics in this regime.

At high disorder correlation ($ \alpha = 3.5 $) we see similar intermediate behavior. For long-range hopping ($\beta = 0.5 $), the spacing distribution again indicates intermediate statistics, with a fitted Brody parameter $ b \approx 0.55 $. However, when both disorder correlation and hopping decay are strong ($ \beta = 3.5 $), the level repulsion becomes pronounced, with $ P(s) $ exhibiting a sharp peak and suppressed small-$ s $ behavior. This regime approaches the Wigner-Dyson limit ($ \beta \approx 1 $), suggesting a delocalized phase with strong level repulsion.

The semi-Poisson and Brody distributions observed in Figure~\ref{fig:spacing} signify intermediate spectral statistics, which in our model can originate from two distinct physical scenarios: (i) a genuine critical or multifractal phase, where eigenstates exhibit scale-invariant properties, or (ii) the coexistence of localized and extended states across different energy windows due to mobility edges. The latter is particularly relevant in our system, as the presence of mobility edges leads to spectral mixing. To distinguish between these scenarios, we rely on energy-resolved diagnostics such as the participation ratio, entanglement entropy, and $\rho$-ratio, which allow us to separate the contributions from different spectral regions.

The number variance $\Sigma^{2}(L)$, shown as insets in each panel, further supports these conclusions. 
For the parameter sets $(\alpha,\beta) = (0.5,0.5)$ and $(0.5,3.5)$, $\Sigma^{2}(L)$ remains closer to the Poisson prediction (albeit with deviations). 
In contrast, for the other two parameter combinations, the behavior of $\Sigma^{2}(L)$ departs from both Poisson and Wigner–Dyson limits, indicating a distinct form of spectral rigidity.

\subsection{Mobility Edge Signatures from Energy-Resolved Observables}
\label{subsec:mobility_edge_PR_Dq}

\begin{figure}[t]
	\centering
	\includegraphics[width=0.45\linewidth]{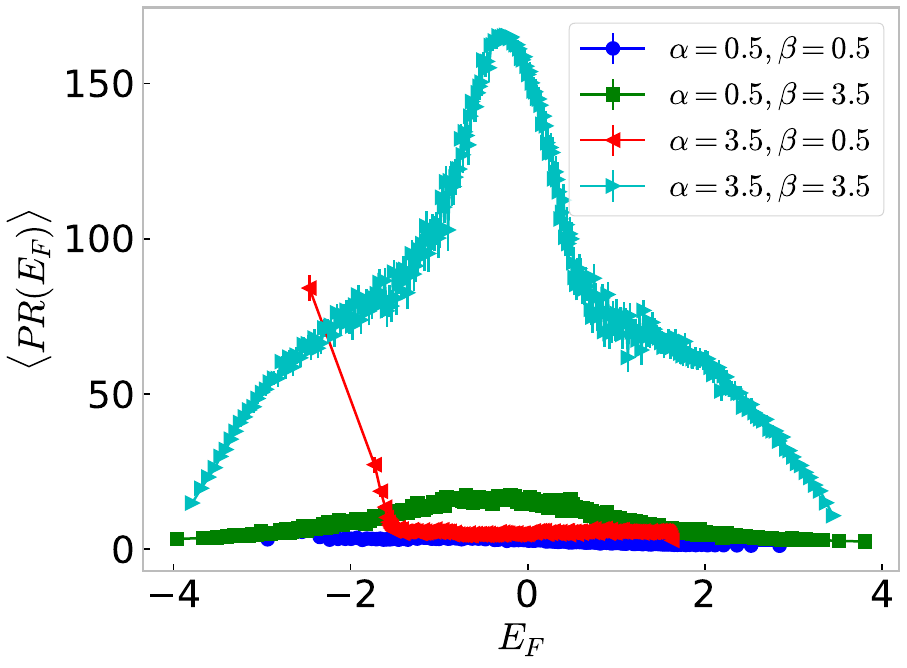}
	\includegraphics[width=0.45\linewidth]{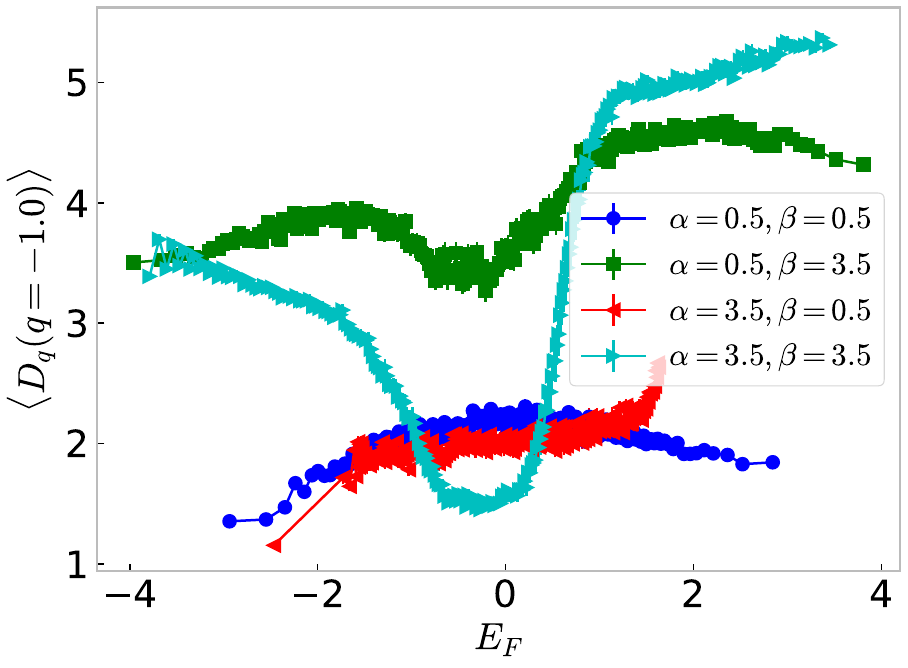}
	\caption{Energy-resolved behavior of (left) average participation ratio $\langle \mathrm{PR} \rangle$ and (right) multifractal dimension $\langle \mathrm{D_q} \rangle (q=-1)$ for selected $(\alpha, \beta)$ pairs. We set the system size $N=256$ and we averaged over $50$ disorder realizations. Standard errors are plotted as a vertical line at each data point. While only data for \(N=256\) are shown for clarity, we have verified that the energy-dependent trends are consistent for system sizes \(N = 128, 384, 512\), indicating that the mobility-edge signatures are robust against finite-size effects.}
	\label{fig:mobility_edge_PR_Dq}
\end{figure}

To identify possible mobility edges—transitions between localized and extended states within the energy spectrum—we examine energy-resolved diagnostics at representative points in the $(\alpha, \beta)$ parameter space. Specifically, we analyze the behavior of the participation ratio $\langle \mathrm{PR} \rangle$ and the multifractal dimension $\langle \mathrm{D_q} \rangle$ as functions of the Fermi energy $E_F$. Figure~\ref{fig:mobility_edge_PR_Dq} presents these quantities for four selected parameter sets: $(\alpha, \beta) = (0.5, 0.5), (0.5, 3.5), (3.5, 0.5)$, and $(3.5, 3.5)$. In all cases, a clear energy dependence emerges.  This behavior supports the existence of a mobility edge, separating a delocalized region from localized states.

\begin{figure*}[t]
	\centering
	\begin{tabular}{cc}
		\includegraphics[width=0.45\textwidth]{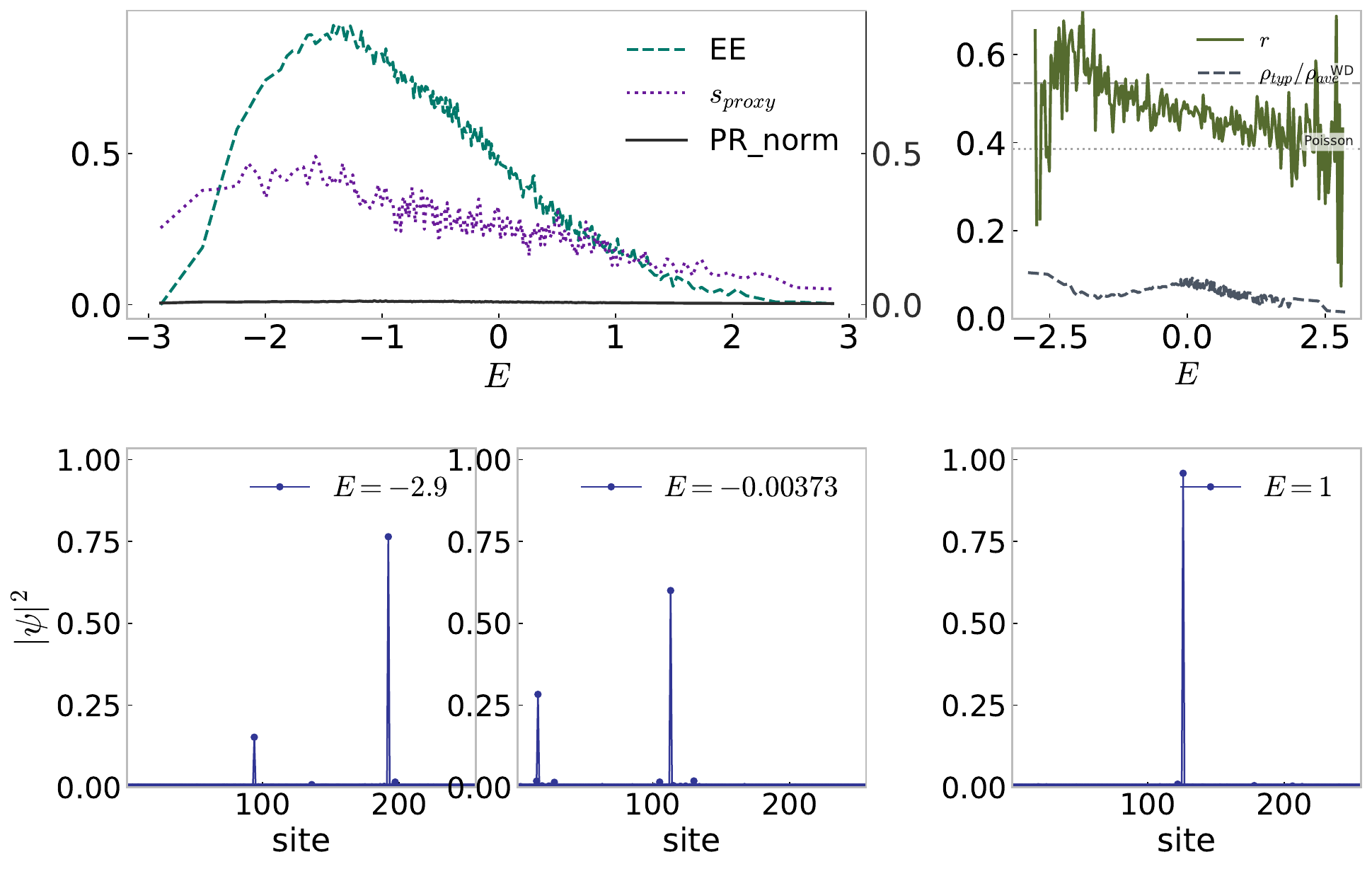} &
		\includegraphics[width=0.45\textwidth]{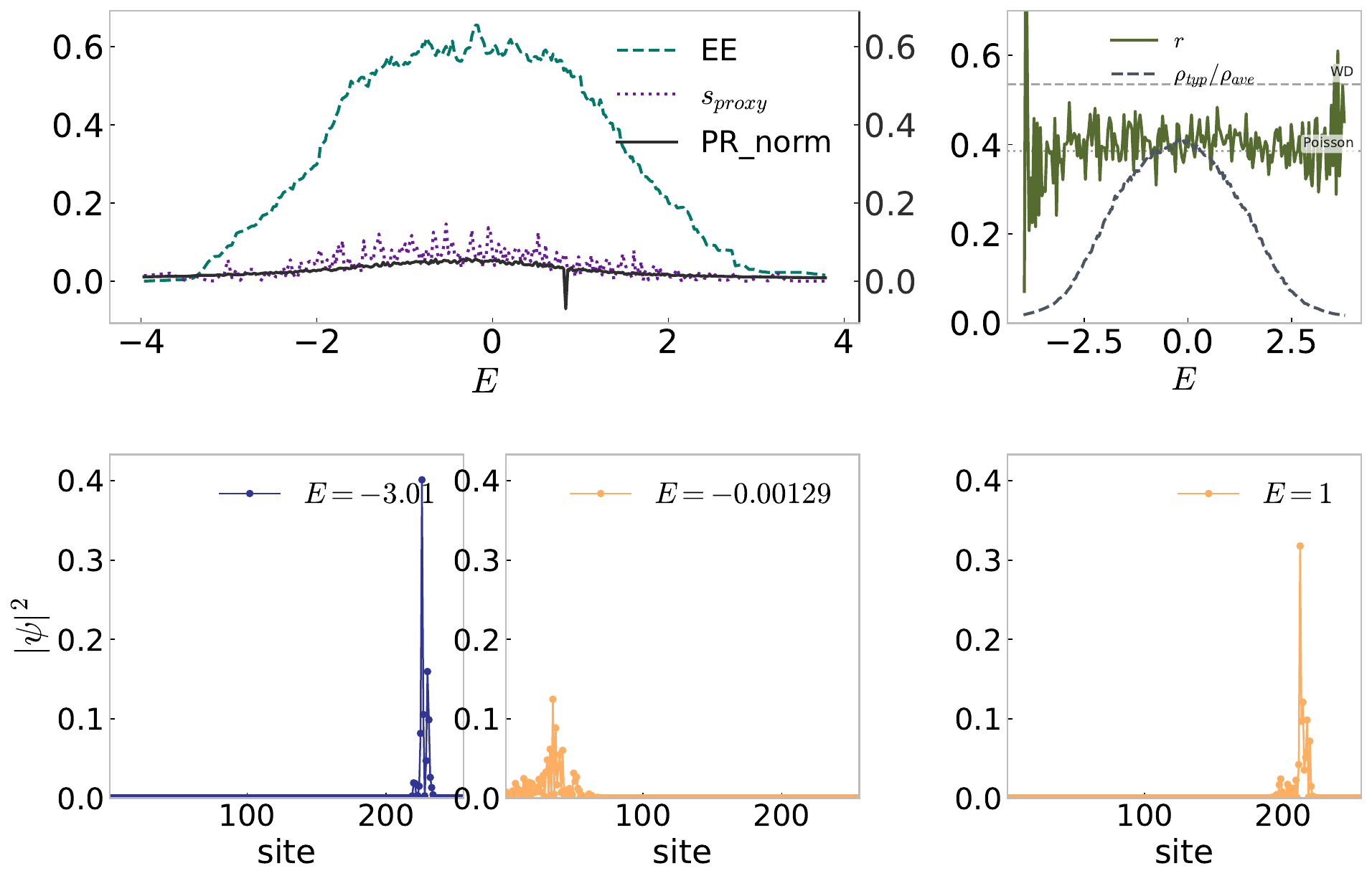} \\
		(a) $(\alpha,\beta)=(0.5,0.5)$ & (b) $(\alpha,\beta)=(0.5,3.5)$ \\[6pt]
		\includegraphics[width=0.45\textwidth]{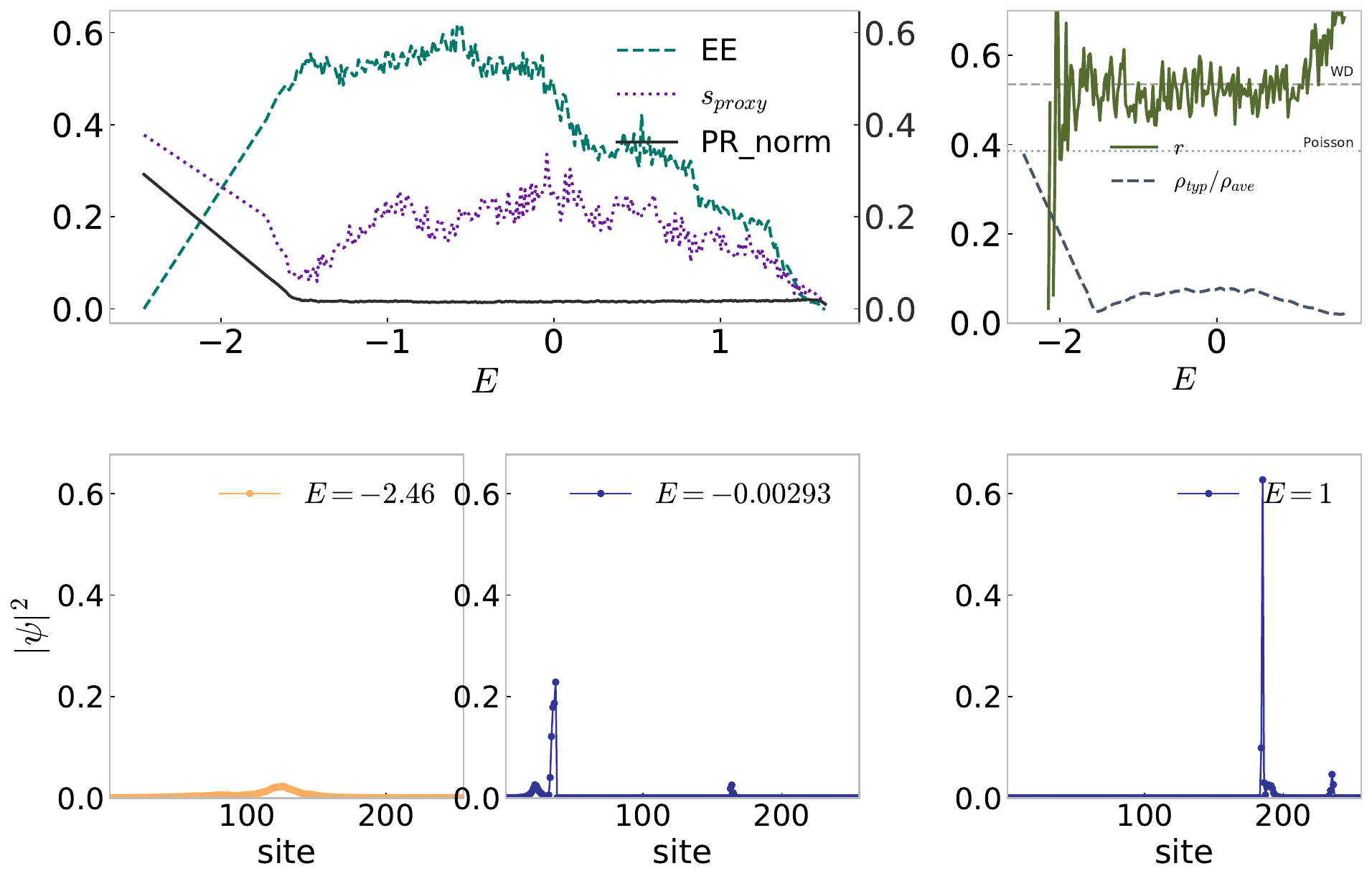} &
		\includegraphics[width=0.45\textwidth]{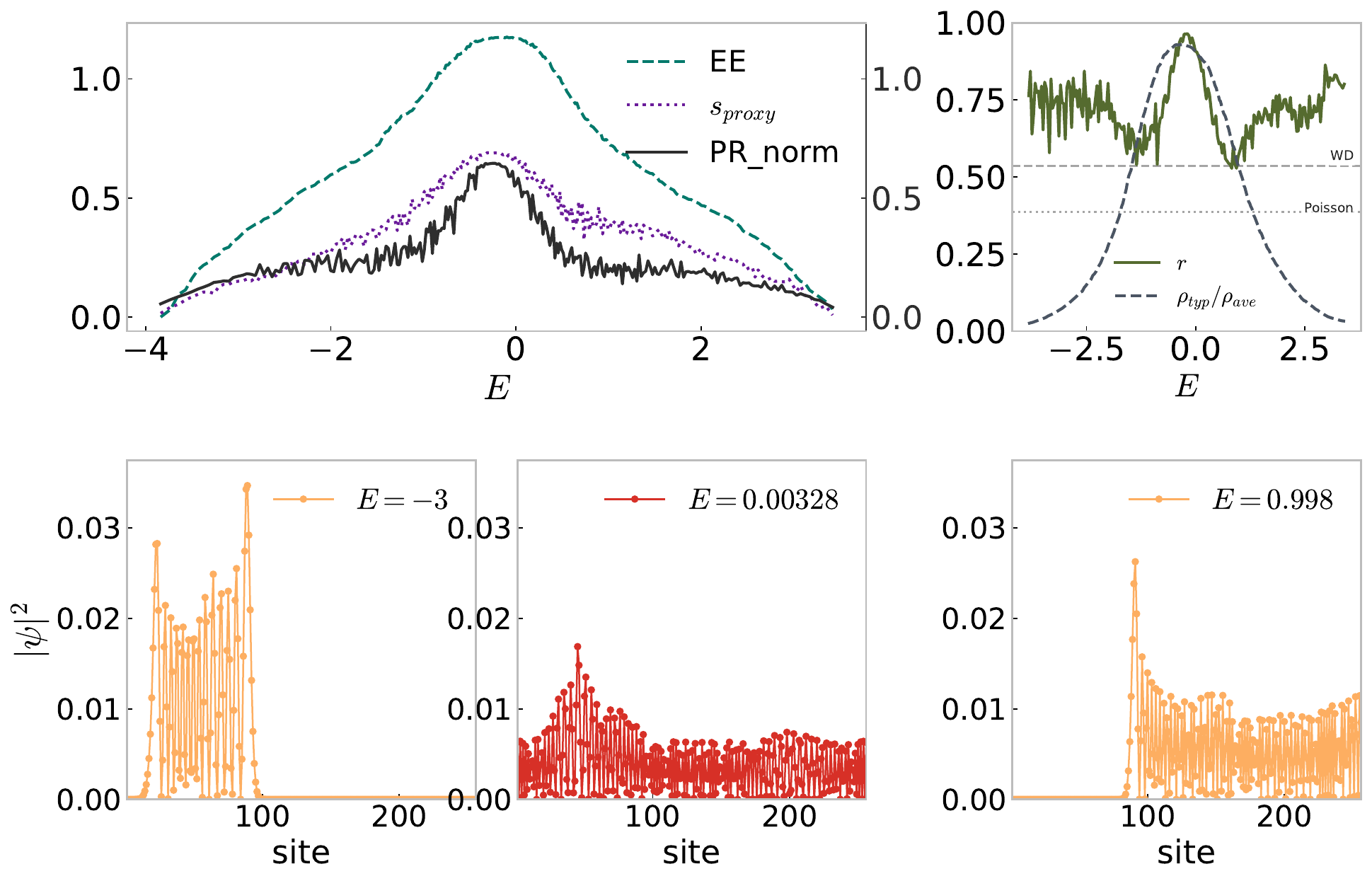} \\
		(c) $(\alpha,\beta)=(3.5,0.5)$ & (d) $(\alpha,\beta)=(3.5,3.5)$
	\end{tabular}
	\caption{Composite local diagnostics for representative parameter points. Each panel contains: Top-left: the normalized participation
		ratio $\mathrm{PR}/N$, plotted together with the
		single-particle entanglement entropy and $s_{\mathrm{proxy}}$, each resolved by energy. Top-right: the binned spectral ratio $r$ and the
		density-of-states ratio $\rho_{\mathrm{typ}}/\rho_{\mathrm{ave}}$,
		both highlighting spectral regions that deviate sharply from the
		expected localized or extended limits. Bottom row: three representative eigenstate intensities $|\psi(i)|^{2}$ taken from energies near the band
		center, mid-band, and band edge.  Each intensity profile is
		colored according to the classifier label associated with its
		energy bin (color scheme consistent with
		Fig.~\ref{fig:classification_maps}). The system size is fixed at $N=256$. For quantities in top panel, each data point represents an average over $100$ realizations, whereas the bottom panels display results from a single representative realization.}
	\label{fig:local_diagnostics}
\end{figure*}

Having established in Sec.~\ref{subsec:level-stats} that the global
level-spacing distribution neither conforms to pure Poisson statistics
(nominally localized) nor to a clean Wigner--Dyson form (nominally
extended), and having shown in
Sec.~\ref{subsec:mobility_edge_PR_Dq} that energy-resolved indicators reveal
well-defined mobility edges, it becomes essential to inspect the full
spectrum using a broader set of local diagnostics.  In particular, we
examine the normalized participation ratio $\mathrm{PR}/N$, the
single-particle entanglement entropy, the smoothness indicator
$s_{\mathrm{proxy}}$, the ratio $\rho_{\mathrm{typ}}/\rho_{\mathrm{ave}}$,
and the spectral ratio $r$ (see Figure~\ref{fig:local_diagnostics}).  These quantities—each sensitive to
distinct aspects of localization, spatial structure, and spectral
regularity—jointly illuminate how different classes of eigenstates
(localized, extended, resonant, and critical) coexist and reorganize
across the band (see below for exact definitions).

\subsection{Classification maps and local diagnostics}
\label{sec:classification_local_diagnostics}

We summarize the parameter-dependent localization behavior using energy--parameter \emph{classification maps}, which assign each narrow energy bin one of four labels (localized, resonant, critical/multifractal, extended) based on a multi-metric criterion. The classifier operates on ensemble-averaged quantities computed in narrow energy bins: the spectral-spacing ratio $r$ (Poisson $\langle r\rangle\!\approx\!0.386$, GOE $\langle r\rangle\!\approx\!0.531$), the typical-to-arithmetic DOS ratio $\rho_{\mathrm{typ}}/\rho_{\mathrm{ave}}$, the normalized participation ratio $\mathrm{PR}/N$, and a spectral-clustering proxy ($s_{proxy}$).  
We use the conservative thresholds
\begin{equation}
	\begin{aligned}
		r_{\mathrm{low}} &= 0.40, &\qquad r_{\mathrm{high}} &= 0.50,\\
		\rho_{\mathrm{lo}} &= 0.15, &\qquad \rho_{\mathrm{hi}} &= 0.60,\\
		\mathrm{PR}_{\mathrm{lo}} &= 0.05, &\qquad \mathrm{PR}_{\mathrm{hi}} &= 0.30,\\
		\text{sproxy}_{\mathrm{th}} &= 0.35 .
	\end{aligned}
	\label{eq:parameters}
\end{equation}

Bins are labeled \emph{extended} only when all spectral, density-of-states, and eigenstate–structure indicators consistently signal delocalization—namely, when the spacing ratio $r$ approaches the GOE value, the typical DOS is comparable to the arithmetic DOS ($\rho_{\mathrm{typ}}/\rho_{\mathrm{ave}}\!\sim\!1$), and the participation ratio is appreciable ($\mathrm{PR}/N$ above threshold), indicating wavefunctions spread over a finite fraction of the system.  
Conversely, a bin is labeled \emph{localized} when these same diagnostics all fall below their low thresholds: Poisson-like $r$, strongly suppressed $\rho_{\mathrm{typ}}$, and small $\mathrm{PR}/N$, together reflecting exponentially localized eigenstates with sparse spectral support.

We identify \emph{resonant} bins as those in which eigenstates display large participation ratios yet retain localized spectral character—specifically, where $\mathrm{PR}/N$ is elevated but $\rho_{\mathrm{typ}}/\rho_{\mathrm{ave}}$ remains small or the clustering proxy (\texttt{sproxy}) indicates poor hybridization across sites.  
These states are understood as being formed from \emph{strong long-range hybridization within a small cluster of sites} that happen to be in resonance due to the algebraic hopping tail, while remaining disconnected from the rest of the lattice.  
Such spatially extended but non-ergodic clusters are reminiscent of the ``necklace'' or ``tail'' states discussed in models with long-range interactions or off-diagonal disorder \cite{Tikhonov_TailStates_2021, Wimmer_Coexistence_2021}.

\begin{figure}[t]
	\centering
	\begin{tabular}{cc}
		\includegraphics[width=0.47\textwidth]{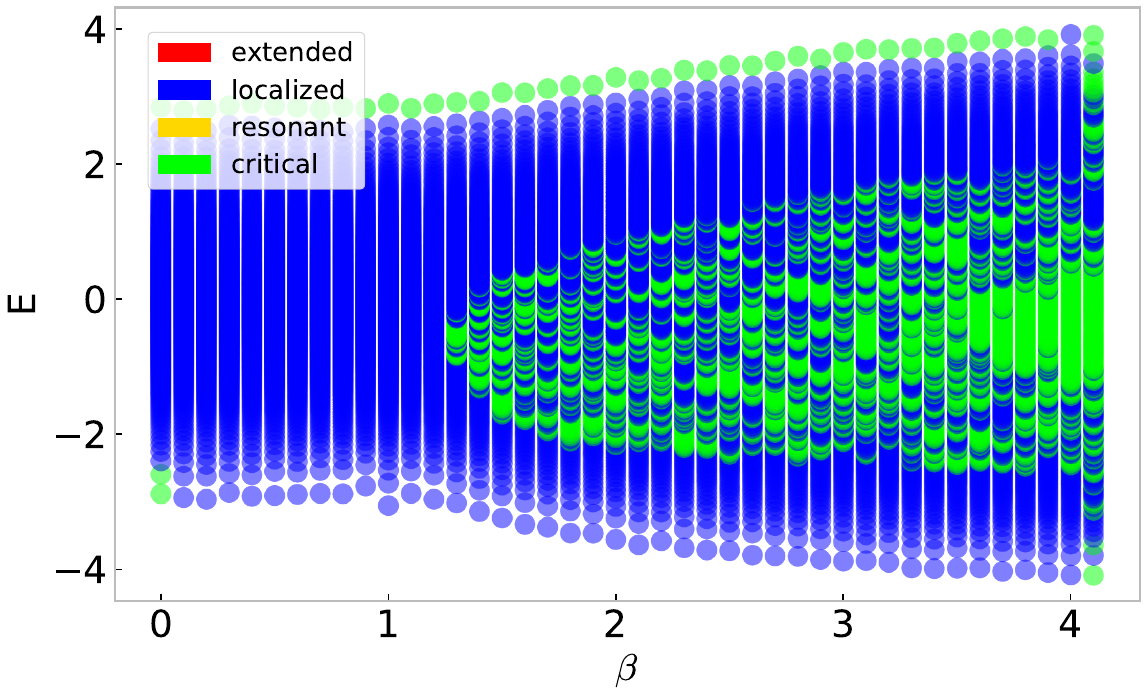} &
		\includegraphics[width=0.47\textwidth]{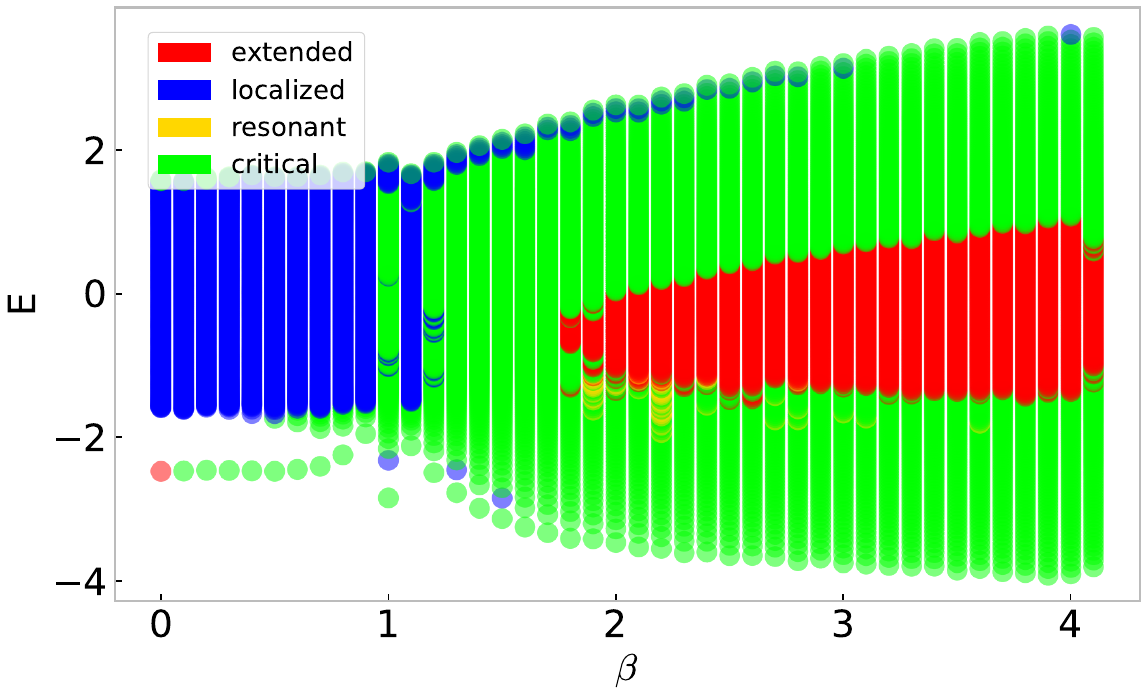} \\
		(a) Fixed $\alpha=0.5$ & (b) Fixed $\alpha=3.5$ \\[6pt]
		\includegraphics[width=0.47\textwidth]{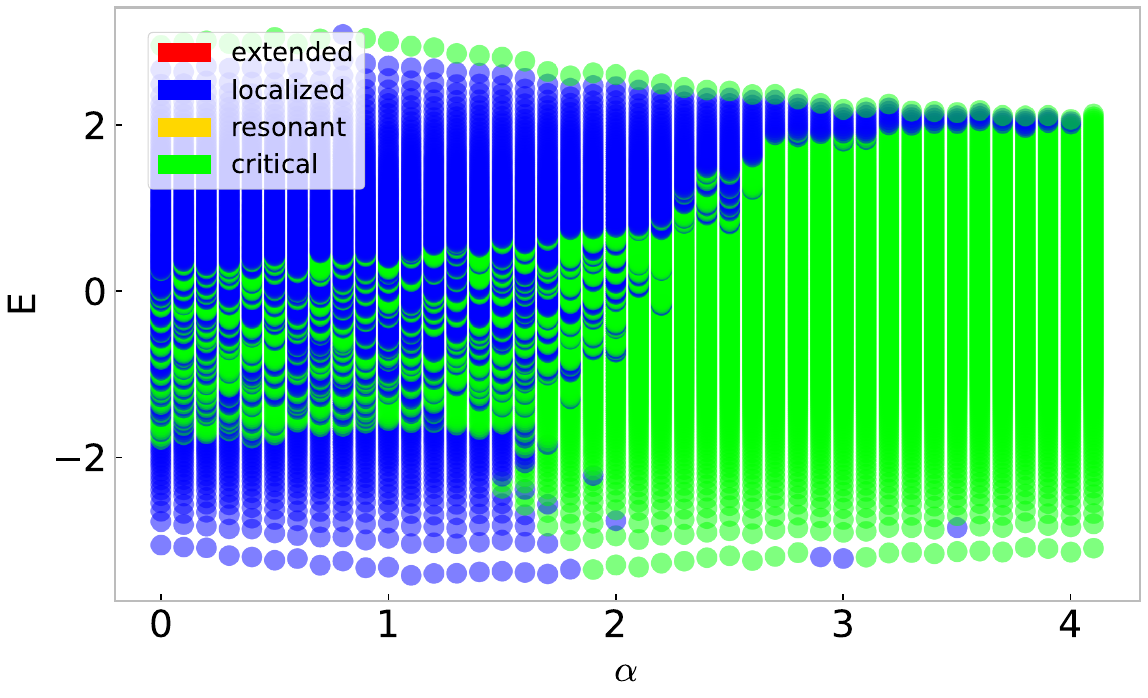} &
		\includegraphics[width=0.47\textwidth]{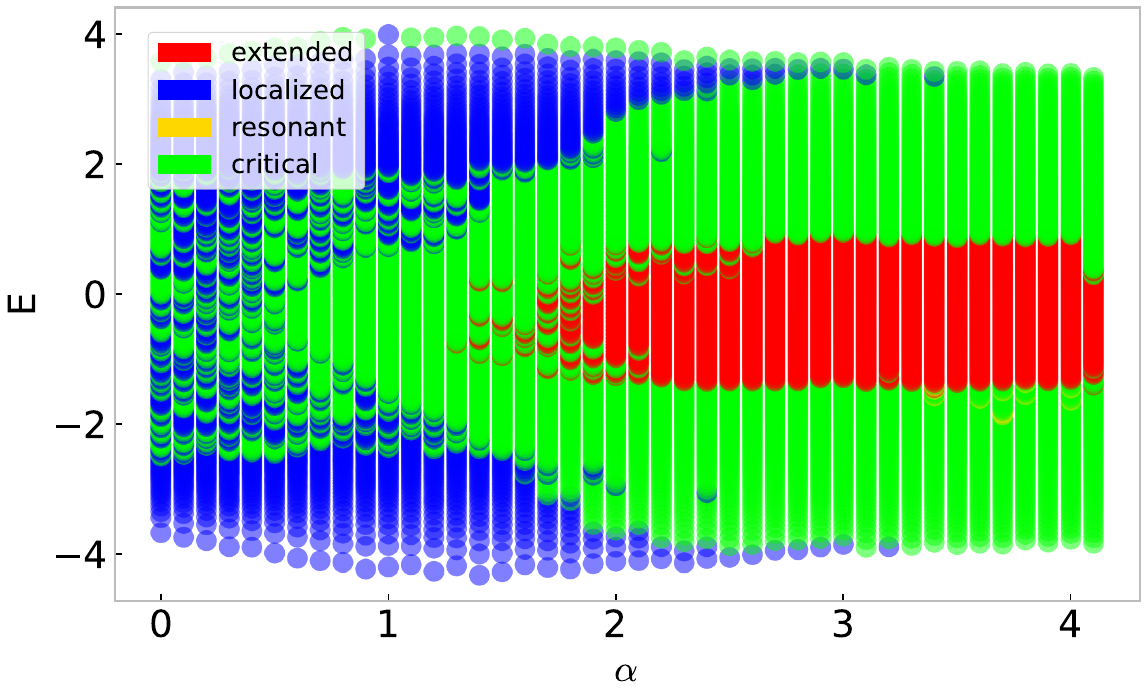} \\
		(c) Fixed $\beta=0.5$ & (d) Fixed $\beta=3.5$ \\
	\end{tabular}
	\caption{Classification maps in the energy--parameter plane. 
		Each panel shows the classifier output at $N=256$, averaged over $100$ disorder realizations.}
	\label{fig:classification_maps}
\end{figure}

Conversely, bins are labeled \emph{critical/multifractal} when different diagnostics point in opposing directions (e.g., moderately enhanced $r$, partial suppression of $\rho_{\mathrm{typ}}$, or intermediate $\mathrm{PR}/N$), reflecting states that are neither localized nor fully extended.  
These states display scale-dependent spreading characteristic of critical or multifractal wavefunctions.  
Physically, critical states exist at mobility edges and exhibit true multifractality and scale invariance, whereas resonant states are \emph{localized on a few, widely separated but strongly hybridized sites}, lacking the scale-invariance characteristic of criticality.
 (see Fig.~\ref{fig:classification_maps})

\paragraph*{Main observations.}  
Across sweeps in either $\alpha$ or $\beta$, we observe a robust sequence: predominantly localized spectra for $\alpha,\beta\!<\!1$; the emergence of critical regions near the band center for $1\!\lesssim\!\alpha,\beta\!\lesssim\!2$; and nucleation of extended states at the band center for $\alpha,\beta\!>\!2$, expanding outward with increasing hybridization.  
Resonant pockets appear in regimes where $\mathrm{PR}/N$ is large but $\rho_{\mathrm{typ}}/\rho_{\mathrm{ave}}$ and $r$ remain localized-like, highlighting the limitation of single-metric interpretations.

\subsection{State-fraction maps in the $\alpha$--$\beta$ plane}
\label{sec:state_fraction_maps}

\begin{figure}[t]
	\centering
	\includegraphics[width=0.95\linewidth]{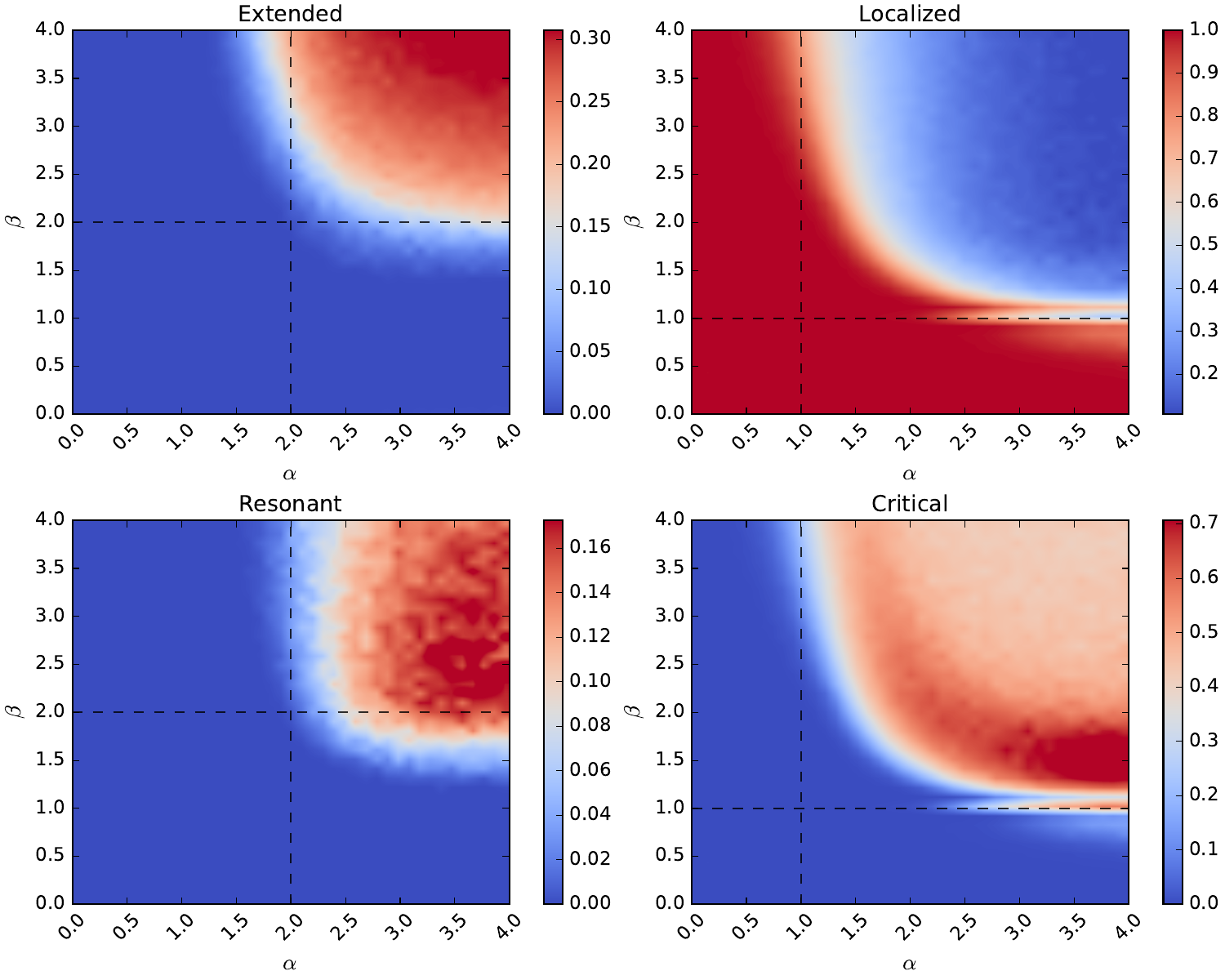}
	\caption{State-weighted fraction maps for $N=512$.  Each panel shows the fraction $f$ of spectral weight classified as (top-left) \emph{extended}, (top-right) \emph{localized}, (bottom-left) \emph{resonant}, and (bottom-right) \emph{critical}.  For each $(\alpha,\beta)$ the spectrum is binned ($n_{\rm bins}=100$) and bins are classified using the composite thresholds on $r$, $\rho_{\rm ratio}$ and $\mathrm{PR}_{\rm norm}$ described in Sec.~\ref{sec:classification_local_diagnostics}; fractions are counts-weighted and disorder-averaged. For each data point, an average over $100$ samples is computed}
	\label{fig:class_fraction_grid}
\end{figure}

Figure~\ref{fig:class_fraction_grid} presents the state-weighted fractions $f_{\mathrm{ext}}$, $f_{\mathrm{loc}}$, $f_{\mathrm{res}}$ and $f_{\mathrm{crit}}$ (panels: extended, localized, resonant, critical, respectively) computed for $N=512$.  For each parameter pair $(\alpha,\beta)$ we subdivide the spectrum into $n_{\rm bins}=100$ energy bins , classify each bin using the composite decision tree described in Sec.~\ref{sec:classification_local_diagnostics} and form state-weighted fractions by summing bin occupancies (counts) per class and normalizing by the total occupied states for that key.  The maps therefore report the fraction of spectral weight assigned to each phenomenological class, averaged over disorder realizations (counts-weighted averaging).

The most salient, reproducible feature is a robust delocalized pocket in the upper-right region of parameter space: states become progressively more extended as both correlation strength ($\alpha$) and the effective short-range normalization parameter ($\beta$) increase (upper-right quadrant).  Conversely, the lower-left region (small $\alpha$ or small $\beta$) is dominated by localized character.  Resonant bins accumulate adjacent to the extended pocket (consistent with large spatial weight but incomplete spectral rigidity), while the critical fraction outlines a curved transition collar that interpolates between localized and extended regimes.  These patterns are qualitatively consistent with the energy-resolved diagnostics shown in Fig.~\ref{fig:local_diagnostics} 

The maps indicate a smooth  crossover surface in $(\alpha,\beta)$ that separates Poisson-like and GOE-like behavior; however, finite-size effects, the energy dependence of the density of states, and the particular threshold choices in the classifier can shift apparent boundaries.  To convert the observed crossover into a bona fide phase transition we must perform finite-size scaling of $f_{\mathrm{ext}}$ and $f_{\mathrm{crit}}$ along representative cuts,  to follow.

\subsection{Finite-size scaling analysis}
\label{sec:fss}

We perform finite-size scaling analyses to extract the critical parameters associated with the transitions observed in the state-class fractions $f$ (localized, resonant, critical, extended). For a chosen observable $f$ (one of the four state fractions) and for a one-dimensional control parameter $t$ (either $\beta$ for fixed $\alpha$ sweeps or $\alpha$ for fixed $\beta$ sweeps), we adopt the two-parameter scaling ansatz in Eq.~\eqref{eq:scaling_ansatz},
where $t_c$ is the critical point, $\gamma_1$ and $\gamma_2$ are finite-size scaling exponents, and $\mathcal{F}$ is an unknown smooth scaling function. Successful data collapse — i.e., all system sizes falling onto a single curve when plotted against the scaling variable $N^{\gamma_2}(t-t_c)$ — provides numerical evidence for a continuous transition.

To determine the optimal collapse parameters $(t_c,\gamma_1,\gamma_2)$, we employ a direct numerical optimization procedure rather than an analytical expansion of $\mathcal{F}$. For each candidate parameter set, we map every data point $(t, f(N,t))$ (for system sizes $N = \{128,256,384,512\}$) to scaled coordinates
\[
x = N^{\gamma_2}(t-t_c), \qquad y = f(N,t)/N^{\gamma_1}.
\]
The collapse quality is quantified using the smoothness-based cost function $\mathcal{C}$ defined in Eq.~\eqref{eq:Csmooth}. Specifically, all $(x_i, y_i)$ pairs are sorted by $x_i$, and the corresponding $y_i$ values form a sequence $\{Q_i\}$. The cost function, Eq. (\ref{eq:Csmooth}), 
measures the normalized total variation of this sequence, penalizing oscillatory deviations from smoothness. A perfect collapse yields a smooth (ideally monotonic) function, minimizing $\mathcal{C}$. The optimal parameters are found by minimizing $\mathcal{C}$ using standard nonlinear optimization routines.

\begin{figure}[htb]
	\centering
	\includegraphics[width=0.95\linewidth]{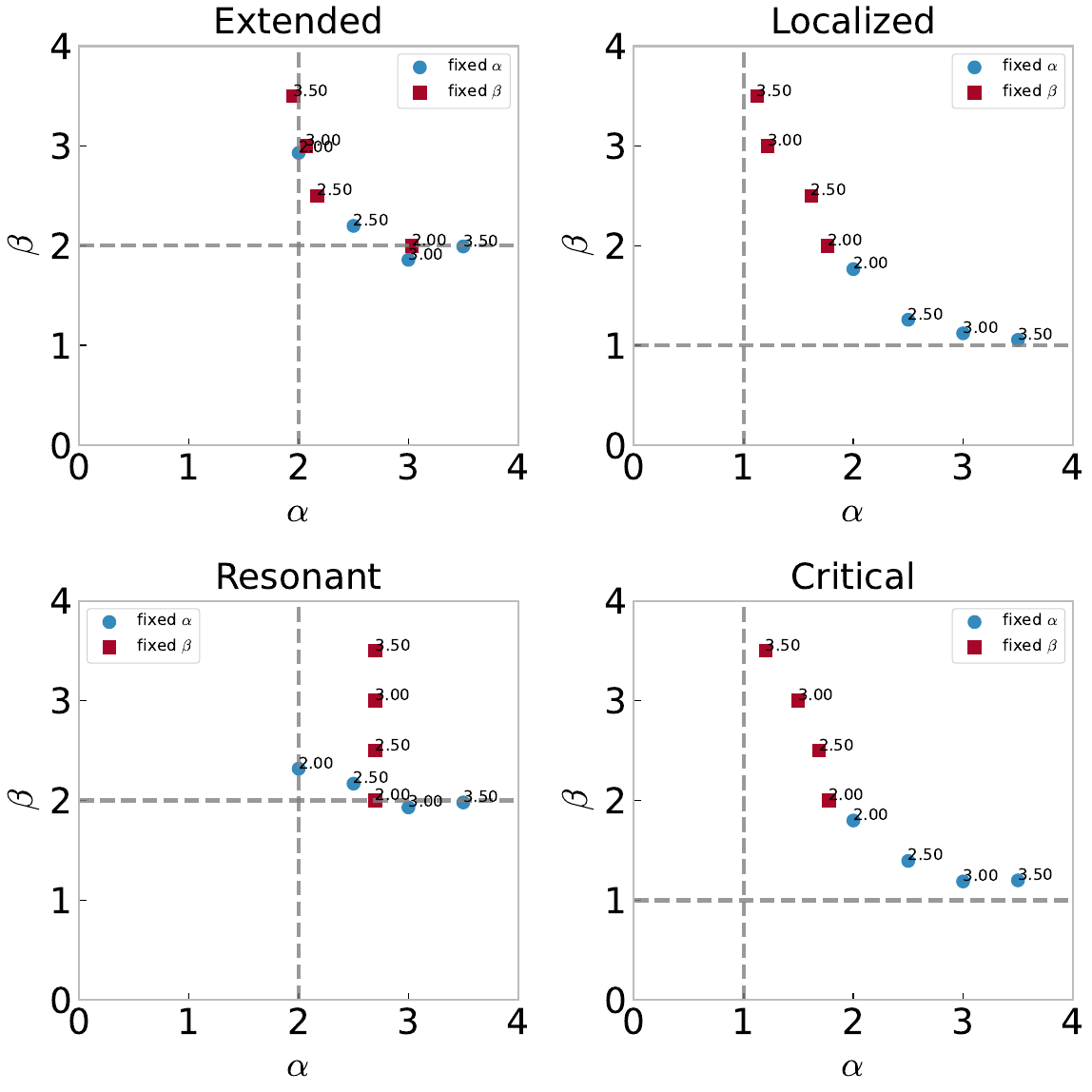}
	\caption{Critical points in the $(\alpha, \beta)$ plane extracted from finite-size collapses of the four state-fractions $f$ (extended, localized, resonant, critical). Circles denote critical points obtained from sweeps in $\beta$ at fixed $\alpha$ (the $\alpha$ value is indicated next to each circle, and the vertical coordinate is the critical $\beta_c$). Squares denote critical points obtained from sweeps in $\alpha$ at fixed $\beta$ (the $\beta$ value is indicated next to each square, and the horizontal coordinate is the critical $\alpha_c$). Thus, each marker corresponds to a transition point along a one-dimensional cut in parameter space.}
	\label{fig:alpha_beta_critical}
\end{figure}

\begin{figure*}[htb]
	\centering
	\begin{tabular}{cc}
		\includegraphics[width=0.45\textwidth]{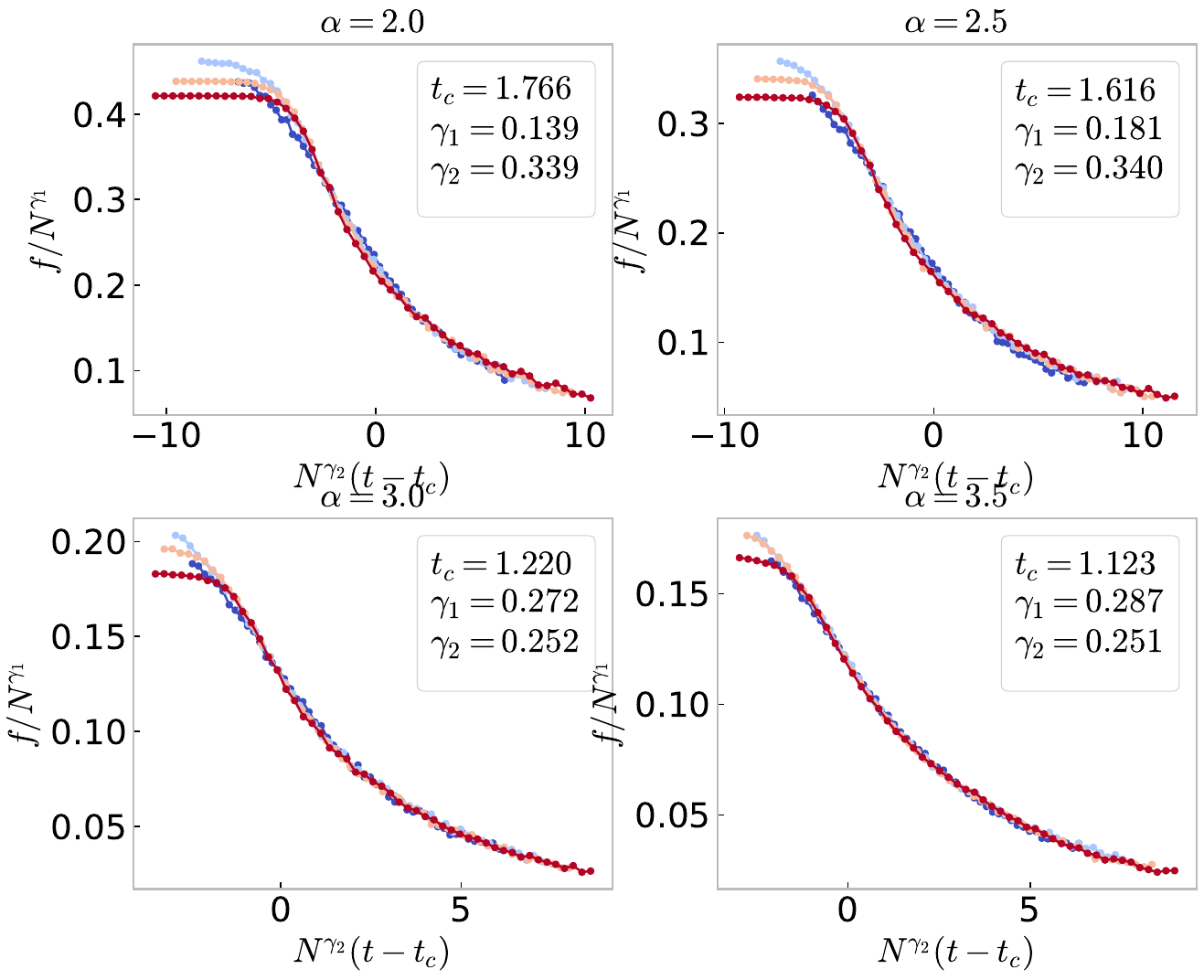} &
		\includegraphics[width=0.45\textwidth]{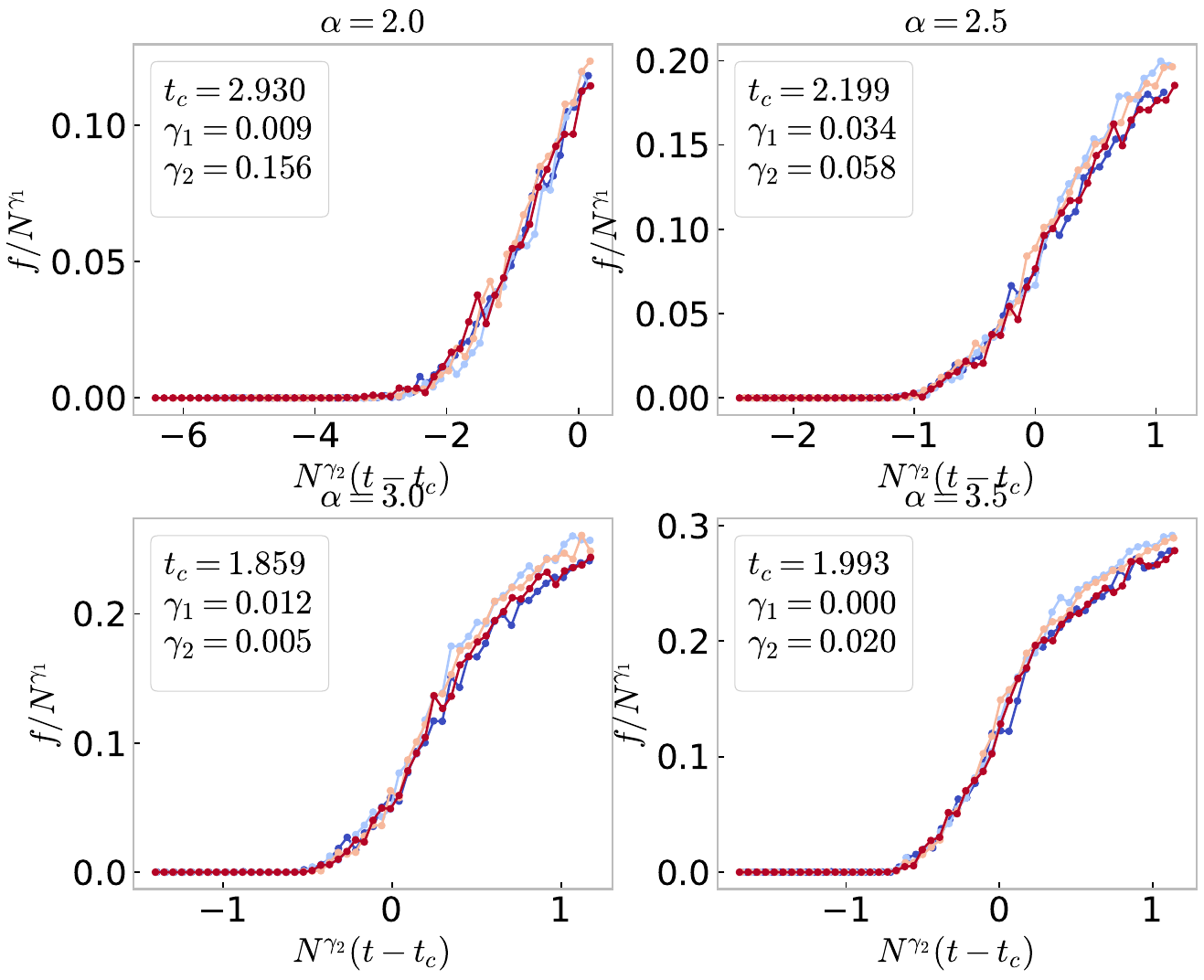} \\
		\includegraphics[width=0.45\textwidth]{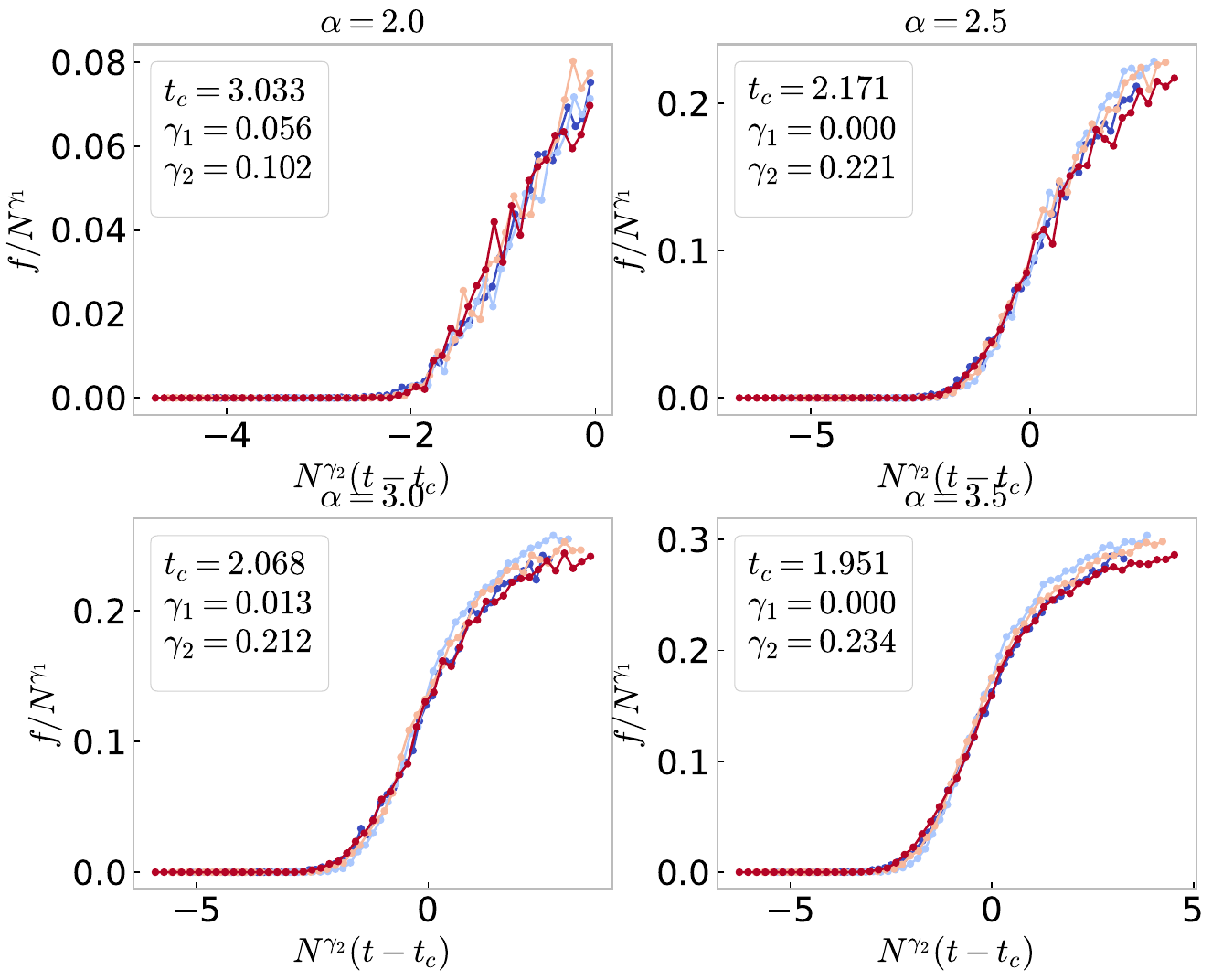} &
		\includegraphics[width=0.45\textwidth]{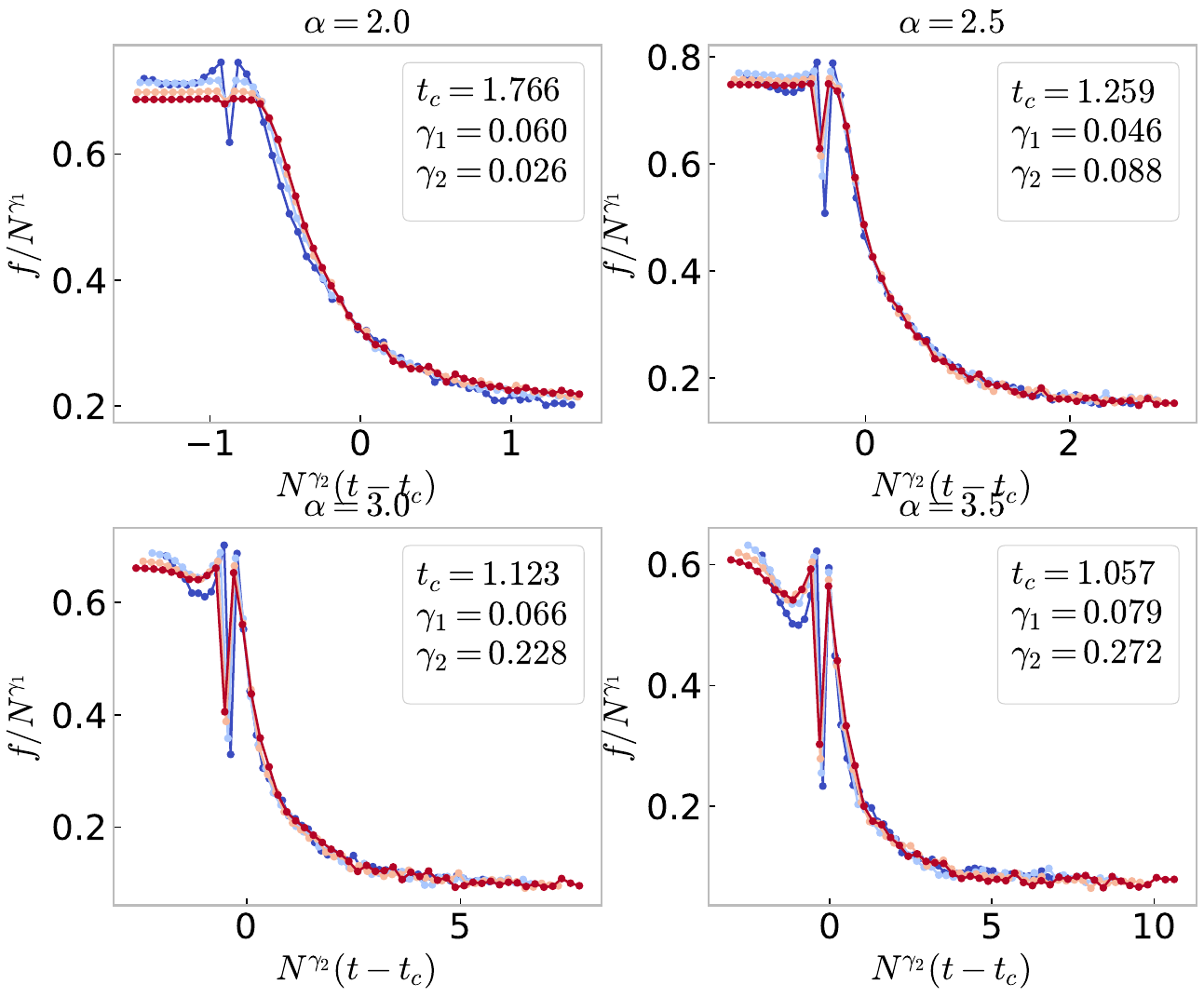}
	\end{tabular}
	\caption{Representative finite-size scaling collapses. For each panel, the inset (legend) reports the optimized parameters $(t_c, \gamma_1, \gamma_2)$. Data correspond to system sizes $N = \{128, 256, 384, 512\}$, shown in blue, cyan, orange, and red, respectively. Each data point represents an average over 100 independent disorder realizations for each system size.}
	\label{fig:example_collapses}
\end{figure*}

For each class $c\in\{\mathrm{localized,resonant,critical,extended}\}$ and for each sweep mode (fixed $\alpha$ or fixed $\beta$), we generate data sets $f(N,t)$ over a dense grid of the tuning parameter $t$. The metrics collapsed in Fig.~\ref{fig:example_collapses} are precisely these four state fractions, as labeled in each panel. Figure~\ref{fig:alpha_beta_critical} displays the extracted critical points in the $(\alpha,\beta)$ plane. The resulting critical points show excellent consistency with the boundaries inferred from Fig.~\ref{fig:class_fraction_grid}.

To illustrate typical collapse quality, Fig.~\ref{fig:example_collapses} presents four representative collapse panels. Each subplot shows several system sizes $N$, the rescaled abscissa $N^{\gamma_2}(t-t_c)$, and the rescaled ordinate $f/N^{\gamma_1}$. The legend reports the optimized parameters $(t_c,\gamma_1,\gamma_2)$ obtained for that fixed value.

\begin{figure}[t]
	\centering
	\includegraphics[width=0.47\textwidth]{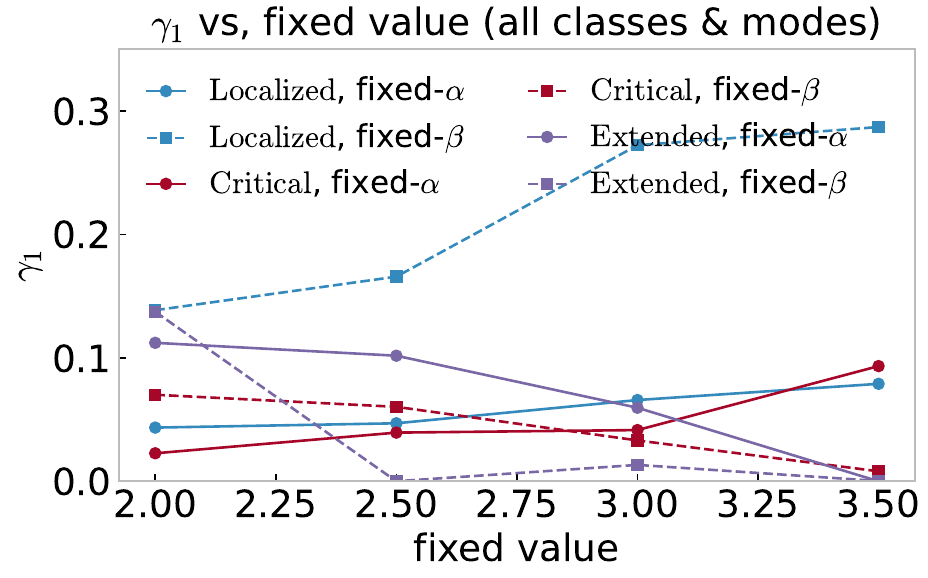}
	\includegraphics[width=0.47\textwidth]{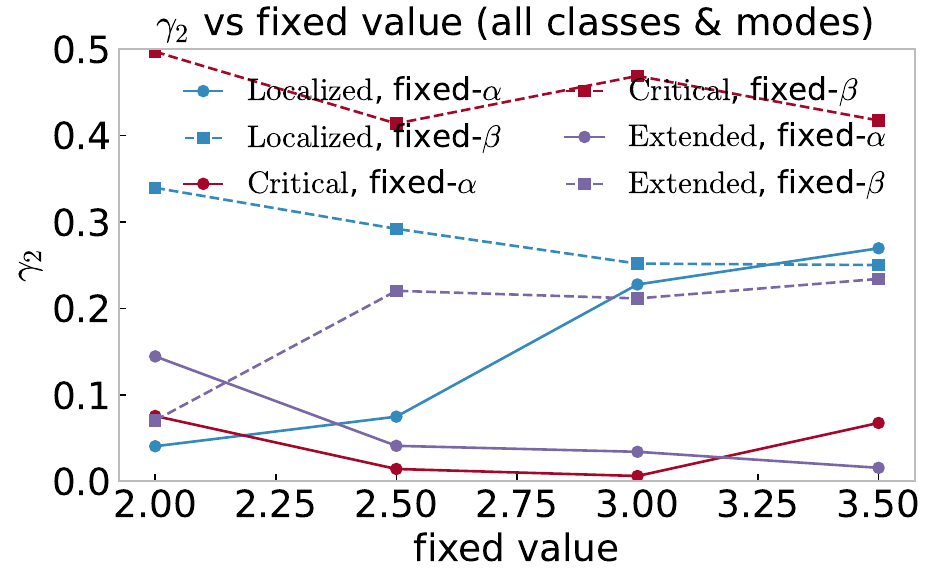}
	\caption{Extracted critical exponents obtained from the smoothness-based
		finite-size scaling analysis. \textit{Left:} Fitted values of $\gamma_1$ for all
		transition lines where a stable collapse is achieved; all data
		satisfy $\gamma_1 < 0.3$.  
		\textit{Right:} Corresponding $\gamma_2$ values, all lying below
		$\gamma_2 < 0.5$.  
		Resonant-regime points are omitted due to inconsistent or
		non-convergent optimization.}
	\label{fig:gamma_exponents}
\end{figure}

A representative set of extracted critical exponents is shown in
Fig.~\ref{fig:gamma_exponents}.  Although one might naively expect
universal exponent values across different transition lines, our data
do not support such a scenario within the accessible system sizes.
Instead, we find that the effective exponents $\gamma_1$ and $\gamma_2$
fall within narrow and well-separated ranges: for all reliably
collapsing cases $\gamma_1 < 0.3$ and $\gamma_2 < 0.5$.  The broad
spread in fitted values likely reflects (i) pronounced finite-size
effects inherent to power-law hopping and correlated disorder,
(ii) the non-universality expected in systems with mobility edges, and
(iii) the sensitivity of the optimization functional to local spectral
features.  Parameter points associated with the resonant regime show
inconsistent or unstable fits; for this reason we do not include them
in the exponent summary.  Overall, the extracted exponent ranges are
robust across observables and provide a consistent characterization of the critical behavior in this model.

\subsection{Machine Learning Analysis of the Emergent Phases}

To complement our statistical and spectral analyzes, we employed machine learning (ML) techniques to investigate whether the phase structure of the system---localized, extended, resonant, and critical---can be inferred directly from the numerical observables without explicit physical input. The objective was twofold: (i) to test if the data possess an intrinsic low-dimensional representation encoding the essential physical organization, and (ii) to assess whether the physically defined quantities $f_j$ (fractions of eigenstates of type $j$) indeed span the natural manifold of the model. If an unsupervised or weakly supervised algorithm autonomously reconstructs the same latent structure as $f_j$, this provides strong evidence that the chosen descriptors are physically meaningful rather than arbitrary parametrizations.

We employ an autoencoder (AE)–based latent representation to extract high-level structure from the spectral data, followed by a supervised readout in latent space.
Each point in parameter space is represented by a feature vector constructed from four spectral observables,
$r$, $\rho_{\mathrm{typ}}/\rho_{\mathrm{ave}}$, $\mathrm{PR}/N$, and an entropy proxy $S_{\mathrm{proxy}}$,
coarse-grained into $40$ uniform energy bins within the window $E\in[-2,2]$.

The AE consists of an encoder–decoder network with a single hidden layer of width $32$ and a latent dimension $d_z=4$.
The network is trained purely unsupervised by minimizing the reconstruction error of the input features; no labels or phase information enter during representation learning.

After unsupervised training, the encoder is frozen and used to map all samples to latent vectors $z$.
Supervision is introduced only at the readout level by training a shallow neural classifier on top of the fixed latent variables.
Class labels are inferred from the dominant phase fraction,
$\arg\max_j f_j(\alpha,\beta)$,
and only parameter points for which all four phase fractions are finite are used for training.
The classifier outputs softmax probabilities for the four regimes—localized, extended, resonant, and critical—which are subsequently mapped back onto the $(\alpha,\beta)$ plane to produce supervised phase diagrams.

Importantly, this supervised step does not modify the learned latent representation.
Instead, it probes and resolves the organization already present in the unsupervised latent space, revealing a clear separation of the four regimes in the encoded representation.

\begin{figure}[t]
\centering	\includegraphics[width=0.95\linewidth]{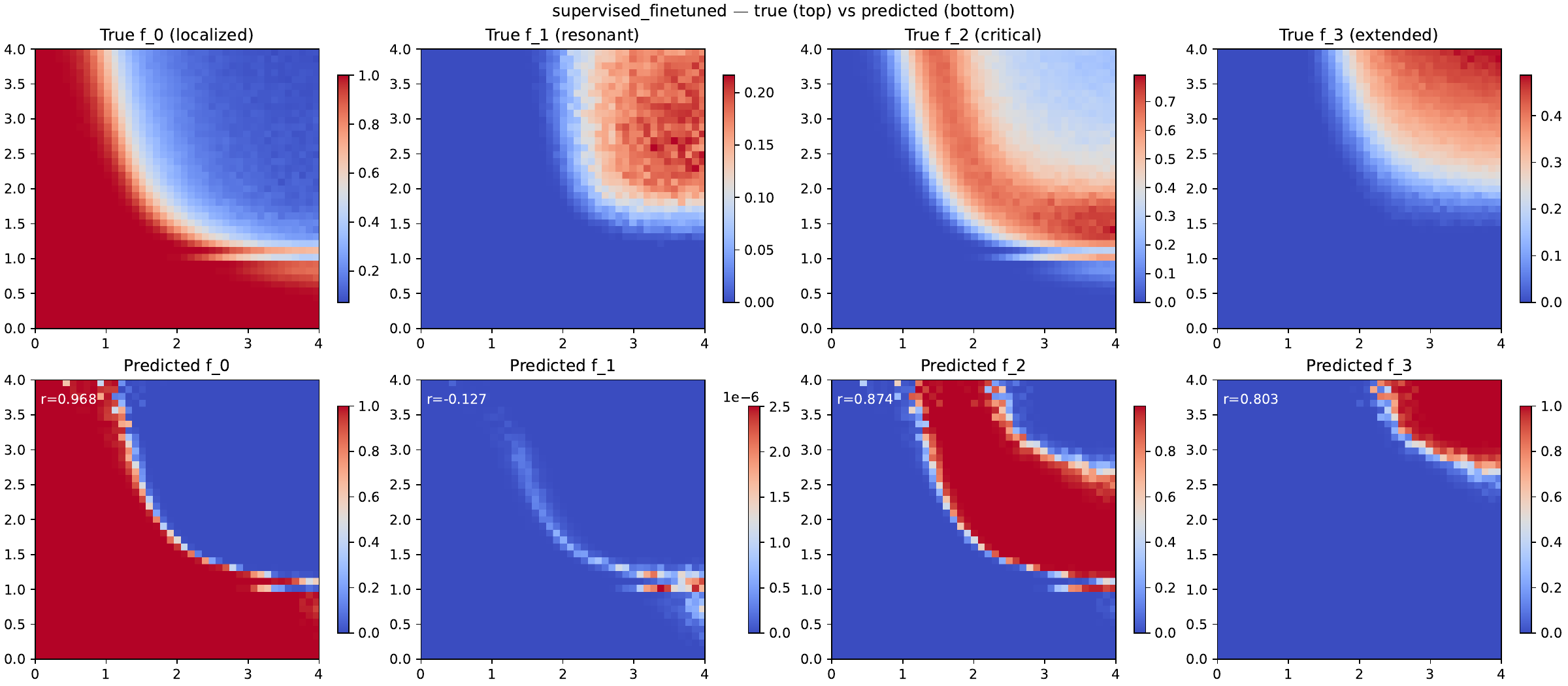}
\caption{(Color online) Comparison of supervised–autoencoder latent coordinates 
$z_i(\alpha,\beta)$ (top row) with physics-based phase fractions $f_j(\alpha,\beta)$ (bottom row) across the four spectral regimes: localized, extended, resonant, and critical. 
The latent representation reproduces the spatial structure of localized, extended, and critical sectors, while the resonant regime shows approximate correspondence due to narrow, irregular spectral windows and strong sample-to-sample fluctuations. 
The Pearson correlation coefficient $r$, shown on each predicted map, quantifies linear agreement between predicted and true values ($r=1$: perfect correlation, $r=0$: no correlation, $r=-1$: anticorrelation), reflecting the predictive quality of the latent representation.}
\label{fig:latent_vs_f}

\end{figure}

After training, the latent projections $z_i(\alpha,\beta)$ organize the data into 
well-separated regions corresponding to the dominant spectral regimes. 
For each latent coordinate, we constructed parameter-space maps and compared them 
with the independently defined phase fractions $f_j(\alpha,\beta)$ (see Fig. \ref{fig:latent_vs_f}). 
For the localized, extended, and critical sectors, the correspondence is striking: 
distinct latent coordinates exhibit a near one-to-one alignment with the respective 
$f_j$ maps, despite the fact that the network was not explicitly instructed to 
reproduce these quantities. In contrast, no single latent coordinate uniquely reproduces the resonant fraction. 
Instead, resonant regions appear as fragmented and partially overlapping structures 
in latent space. This behavior reflects the intrinsically narrow, irregular, and 
sample-dependent nature of resonant states, already evident in the physical 
diagnostics. The absence of a clean one-dimensional latent parametrization therefore 
does not signal a failure of the model, but rather confirms that the resonant regime 
does not constitute a robust manifold in the same sense as the localized, extended, 
or critical phases.

The convergence between the ML-derived latent variables and the physical phase fractions has direct physical significance. First, it confirms that the chosen $f_j$ constitute a faithful parametrization of the true physical manifold governing the localization--delocalization crossover. In other words, the autoencoder---trained without explicit knowledge of the Hamiltonian---discovers the same low-dimensional structure encoded in the theoretically defined $f_j$. This implies that the statistical features provided to the network inherently contain all necessary physical information, validating that the $f_j$ are not arbitrary constructs but emergent quantities naturally arising from the data.

Second, the success of the supervised AE demonstrates the capability of machine learning as a diagnostic and verification tool in condensed matter physics. Here, ML functions as an independent, data-driven probe: it confirms the phase organization and provides an unbiased mapping of the parameter space. The machine learning analysis serves both as a verification and as a discovery mechanism. This result not only strengthens the validity of the analytical and numerical findings but also highlights the potential of hybrid physics--ML methodologies for identifying order parameters and phase boundaries in complex quantum systems.

\section{Conclusion and Outlook}
\label{sec:conclusion}
In this work we have carried out a comprehensive investigation of a one-dimensional tight-binding model featuring the simultaneous presence of long-range correlated onsite potentials (controlled by the exponent $\alpha$) and algebraically decaying long-range hopping amplitudes (parameterized by $\beta$). Using exact diagonalization for system sizes up to $N=512$, we combined level statistics, energy-resolved localization measures, phase-characterization functions $f_i$, finite-size scaling analyzes, and machine-learning–based classification to produce a coherent characterization of the phase structure in the $(\alpha,\beta)$ plane.

Our results demonstrate that the interplay between correlation in the disorder potential and power-law hopping leads to robust mobility edges across significant regions of the phase diagram. Energy-resolved participation ratios, entanglement measures, and the behavior of $r$-statistics collectively reveal spectral windows in which eigenstates transition as a function of energy. 

The presence of robust mobility edges and intermediate phases in our model stems from the interplay between two energy-dependent mechanisms. The correlated disorder creates a spatially smooth potential landscape that reduces scattering, while the long-range hopping enables non-local resonances. Their competition leads to an energy-selective delocalization: states near the band center experience an effectively weaker disorder and become extended, whereas states near the band edges remain localized. This spectral inhomogeneity, absent in models with only one of the two ingredients, gives rise to the observed mobility edges and intermediate critical and resonant regimes.

The four phase-characterization functions $f_1$–$f_4$, synthesized from complementary diagnostics, provide a compact and operationally effective parameterization of the emergent phases. Finite-size scaling, aided by an explicit smoothness-enforcing cost function, yields consistent estimates for transition points and reveals approximate scaling exponents, though we emphasize that uncertainty remains substantial due to limited accessible system sizes.

A supervised autoencoder was trained on local spectral features, showing that its latent variables faithfully reproduce the structure of the $f_i$-maps. This agreement provides an independent validation of our physics-driven diagnostics and indicates that the set of handcrafted observables captures the essential phenomenology of the localization transitions.

Beyond the results presented here, several substantive questions remain open. 
First, the sharp phase boundaries inferred from the $f_i$ maps suggest underlying 
analytical structures that are not yet understood; developing perturbative or 
continuum approximations—particularly in the limits of large or small $\beta$—may 
clarify the origin of these transitions. 
Second, the multifractal properties observed in the mixed and critical regimes 
warrant deeper analysis, especially regarding how long-range hopping modifies 
standard multifractal scaling relations. 
Also, extending the present framework to interacting systems remains a 
compelling direction: the combined effects of long-range correlated disorder 
and algebraic hopping may produce unconventional many-body mobility edges or 
alter the stability of many-body localization, questions that are presently 
unresolved.
In addition, an intriguing open question is whether the level-spacing ratio $r$ takes a universal critical value at the mobility edges in our one-dimensional model, analogous to the universal critical amplitude $r_c \approx 0.513(5)$ found at the three-dimensional Anderson transition\cite{PhysRevA.92.063621}. Given that our model exhibits multiple mobility edges between different spectral regimes (localized, critical, resonant, extended), a systematic investigation of the critical $r$ value and its potential universality across the $(\alpha,\beta)$ plane represents a valuable direction for future research. Finally, we note that alternative diagnostics such as the distribution of extremal eigenvalues of the reduced density matrix\cite{PhysRevB.106.L060201,PhysRevB.106.L060201} have been shown to provide sharp probes of localization transitions with reduced finite-size effects. Applying such methods to our model could yield even more precise estimates of critical points and mobility edges, and we leave this interesting direction for future work.

Overall, our study provides a unified framework to analyze and interpret localization phenomena in long-range, correlated disordered systems and opens up several pathways for future developments in both analytical and numerical fronts.

\acknowledgements

This work was supported by the Iran National Science Foundation (INSF).

\section*{Author Contributions}
M.P. conceived the study, designed the theoretical model, performed the numerical simulations and data analysis, developed and implemented the machine-learning framework, interpreted the results, and wrote the manuscript.

\section*{Competing interests}
The author declares no competing interests.

\section*{Data availability} 
The data supporting the findings of this study are available from the corresponding author upon reasonable request.

\bibliographystyle{apsrev4-2.bst}
\bibliography{reference.bib}
\end{document}